\documentclass[preprint,12pt]{elsarticle}

\usepackage{amssymb}
\usepackage{amsmath}
\usepackage[subrefformat=parens,labelformat=parens]{subfig}
\usepackage{float}
\usepackage{setspace}

\newcommand{\beq}{\begin{equation}}
\newcommand{\eeq}{\end{equation}}

\newcommand{\nene}[1]{#1\pm 1}

\newcommand{\mean}[1]{\overline{#1}}

\journal{Journal of Computational Physics}

\begin{document}

\begin{frontmatter}



\title{Hybrid approaches for multiple-species stochastic reaction-diffusion models}


\author[bu,mit]{Fabian Spill}
\author[ucl]{Pilar Guerrero}
\author[crm,auto]{Tomas Alarcon}
\author[wcmb]{Philip K. Maini}
\author[wcmb,cbg]{Helen Byrne}

\address[bu]{Department of Biomedical Engineering, Boston University, 44 Cummington Street, Boston MA 02215, USA}
\address[mit]{Department of Mechanical Engineering, Massachusetts Institute of Technology,\\ 77 Massachusetts Avenue, Cambridge, MA 02139, USA\\
fspill@bu.edu, fspill@mit.edu}
\address[ucl]{Department of Mathematics, University College London, Gower Street, London WC1E 6BT, UK}
\address[crm]{Centre de Recerca Matematica, Campus de Bellaterra, Edifici C, 08193 Bellaterra, (Barcelona), Spain}
\address[auto]{Departament de Matem\`atiques, Universitat Aton\`oma de Barcelona, 08193 Bellaterra (Barcelona), Spain}
\address[wcmb]{Wolfson Centre for Mathematical Biology, Mathematical Institute, University of Oxford, Oxford OX2 6GG, UK}
\address[cbg]{Computational Biology Group, Department of Computer Science, University of Oxford, Oxford OX1 3QD, UK}

\begin{abstract}
Reaction-diffusion models are used to describe systems in fields as diverse as physics, chemistry, ecology and biology. The fundamental quantities in such models are individual entities such as atoms and molecules, bacteria, cells or animals, which move and/or react in a stochastic manner. If the number of entities is large, accounting for each individual is inefficient, and often partial differential equation (PDE) models are used in which the stochastic behaviour of individuals is replaced by a description of the averaged, or mean behaviour of the system. In some situations the number of individuals is large in certain regions and small in others. In such cases, a stochastic model may be inefficient in one region, and a PDE model inaccurate in another. To overcome this problem, we develop a scheme which couples a stochastic reaction-diffusion system in one part of the domain with its mean field analogue, i.e. a discretised PDE model, in the other part of the domain. The interface in between the two domains occupies exactly one lattice site and is chosen such that the mean field description is still accurate there. This way errors due to the flux between the domains are small. Our scheme can account for multiple dynamic interfaces separating multiple stochastic and deterministic domains, and the coupling between the domains conserves the total number of particles. The method preserves stochastic features such as extinction not observable in the mean field description, and is significantly faster to simulate on a computer than the pure stochastic model.
\end{abstract}


\end{frontmatter}

\section{Introduction}\label{sec:intro}

Random effects due to finite numbers of components are ubiquitous in reaction-diffusion systems. Within this context, much research
has been done, for instance, on the robustness and response to noise in gene regulatory networks. The study of such systems, as well as
other examples such as chemical pattern-forming reaction-diffusion systems, has also revealed that an
accurate description of their dynamics may require inclusion of the effects of spatially-inhomogeneous distributions of molecules. The usual framework to analyse such situations is that of stochastic reaction-diffusion models.            

A popular approach for studying stochastic reaction-diffusion models involves decomposing the domain occupied by the system into small compartments or voxels, and counting the number of molecules of each species in each voxel \cite{stundzia1996stochastic,bernstein2005simulating}. Chemical reactions between species are treated locally (i.e. within each voxel), their rates being determined by the local abundance of each constituent species. Molecular transport between compartments is usually modelled by simple diffusion. 

This compartmental approach poses a problem: numerical (Monte Carlo) techniques such as the Gillespie stochastic simulation algorithm (SSA)\cite{gillespie1976general,gillespie1977exact} or the $\tau-$leap method \cite{gillespie2001approximate,cao2006efficient} are inefficient since they scale poorly with the number of reaction channels, which is proportional to the number of compartments. 

In order to overcome issues regarding the efficiency of direct simulation methods, several strategies have been proposed. For example, the Next Reaction Method (NRM) proposed by Gibson \& Bruck \cite{gibson2000} is an exact method in the same sense as the SSA, i.e. the sample paths generated are exact realisations of the solution of the corresponding Master Equation. However, unlike the SSA, the computing time grows logarithmically with the number of reaction channels. This is accomplished by (i) recycling the random numbers generated in previous time steps, so at each time step only one new random number must be generated, and (ii) organising the reaction channels in a queue (more specifically, a tree) with events ordered in ascending order of waiting time. In this way, it is possible to determine which channel will fire next when by looking at the root of the tree. The Next Sub-Volume Method (NSVM) is a modification of the NRM which accounts for reaction-diffusion systems \cite{elf2004}. It deals with the stiffness problem associated to the fact that the transition rates associated with diffusion are proportional to $h^{-2}$, where $h$ is the compartment length.   

An alternative set of methods are hybrid methods. Often a stochastic description is only needed in a certain region of the domain. Elsewhere the number of components is large enough that a mean-field description is reasonable. A paradigmatic example of this situation is front propagation in reaction-diffusion systems \cite{breuer1994fluctuation,breuer1995macroscopic,moro2004}. In systems such as the stochastic Fisher-Kolmogorov model, the number of particles ahead of the front is small and, therefore, fluctuations need to be taken into account. By contrast, behind the front, the number of particles fluctuates about the carrying capacity of the model. In these conditions, simulating the system using the compartmental approach and a direct (Gillespie) simulation method is inefficient and it is natural to propose a hybrid approach where the region behind the front is modelled as a Fisher-Kolmogorov PDE, the region ahead of the front is treated as a stochastic process, and appropriate matching conditions are applied in the intermediate region. 

This idea, and variations thereupon, have been implemented for several different methods and several systems. Its first incarnations consisted of hybrid models for pure diffusion \cite{flegg2012,flekkoy2001coupling,franz2013}. In \cite{franz2013} uni-molecular chemical reactions such as chemoabsorption are also considered. In all these methods the boundary or overlapping region between the two regimes is considered fixed.

A further step forward towards algorithms that can cope with more complex situations has recently been proposed by Hellander et al. \cite{hellander2012}. Based on previous work on a method to simulate stochastic reaction-diffusion systems on unstructured domains \cite{engblom2009}, they have proposed a method where, at each voxel, the different species are divided into two classes, namely, mesoscopic (i.e. well-mixed within the voxel and with dynamics determined by the corresponding Master Equation) and microscopic. Concerning the latter, they are assumed to be subject to off-lattice reaction-diffusion dynamic, modelled in terms of the Green's function method for the corresponding Smoluchowski equation proposed by Van Zon and ten Wolde \cite{vanzon2005prl,vanzon2005jcp}. The coupling between mesoscopic and microscopic degrees of freedom within each voxel is accomplished via a splitting scheme. Similar hybrid algorithms are discussed in \cite{alexander2002algorithm,alexander2005algorithm,li2010spatially,li2012spatially,flegg2013convergence,
robinson2014adaptive,sanft2015constant,hepp2015adaptive,yates2015pseudo}.

Moro \cite{moro2004} proposed a similar method for simulating stochastic reaction-diffusion systems with propagating fronts in which a macroscopic PDE is coupled with a mesoscopic Master Equation. The two descriptions hold in different sub-domains and are coupled across a moving boundary, using a method which balances the fluxes between the sub-domains on average only. 

The hybrid algorithm presented in this paper also couples a mesoscopic description of a stochastic reaction-diffusion system, modelled by an 
on-lattice, reaction-diffusion Master Equation (RDME), with a 
reaction-diffusion system, which is obtained from the mean field equations associated with the stochastic model. We remark that whilst our mean-field reaction-diffusion system converges to a reaction-diffusion PDE in the continuum limit, the RDME does not converge to the 
associated off-lattice (Doi or Schmolokowski) reaction-diffusion models \cite{isaacson2013convergent,hellander2015} in spatial dimensions larger than 
2. In light of this, our method should be viewed as the hybridisation of the RDME with its associated on-lattice, mean-field reaction 
-diffusion limit.

Our method extends the work of \cite{moro2004} to systems without propagating fronts. Furthermore, the 
interface condition used here preserves the total amount of particles at all times in every single simulation. The interface is also chosen such that 
the mean field description is still valid at the interface region, and each interface is only one compartment in size. This way, errors in the flux 
between the two domains are negligible. We allow multiple interfaces so the stochastic and deterministic regions need not be connected, and interfaces 
may be dynamic in space and time. For multi-species models, different species may exist in different stochastic and deterministic sub-domains. We test 
our algorithm on two classical systems, the Fisher-Kolmogorov equation and a spatial Lotka-Volterra system. The former serves as the simplest example 
of a system 
with a propagating front, and in our hybrid model a single interface separates the stochastic region at the wave front from the deterministic region behind the wave front. The spatially resolved Lotka-Volterra system serves as a test case for a multi-species reaction-diffusion system, where the hybrid model can, in general, have multiple species-specific interfaces.

The paper is organised as follows: In section \ref{sec:stochasticAndPDE} we introduce notations and conventions for stochastic and deterministic reaction-diffusion systems. The methodology for a generic hybrid model with an arbitrary number of species is presented in section \ref{sec:generalAlgorithm} and our algorithm is applied to the Fisher-Kolmogorov equation in section \ref{sec:Fisher}. Finally, the spatially resolved Lotka-Volterra model is investigated in section \ref{sec:LotkaVolterra}.

\section{Stochastic and Deterministic Reaction-Diffusion Systems}\label{sec:stochasticAndPDE}
\subsection{Stochastic Reaction-Diffusion Systems}

We consider a system of $s_{max}$ species defined on a regular lattice. Each species comprises individuals that can migrate to neighbouring lattice sites, or react locally with entities of the same or other species. For simplicity, we describe the system in one spatial dimension, where the lattice index is $k=1,\dots,k_{max}$, but the generalisation to higher dimensions is straight-forward. We let $N_s(k,t)$, $s=1,\dots,s_{max}$ denote the number of individuals of species $s$ in box $k$ at time $t$, and let $h$ denote the lattice constant, so that $L=hk_{max}$ is the domain size. If $N_s(k,t)$ is Markovian, then the time evolution of $N_s(k,t)$ is governed by a master equation of the form
\begin{align}
\label{eq:generalMasterEquation}
 \frac{dP(N,t)}{dt} &= \sum_{\tilde N}(\mathcal{T}_{N| {\tilde N}}P(\tilde{N},t) - \mathcal{T}_{\tilde{N}|N}P(N,t)).
\end{align}
Here, $N$ denotes a generic state specified by the number of individuals $N_s(k,t)$ of all species $s$ in any compartment $k$, and likewise ${\tilde N}$, so the sum is understood to be over all states defined by ${\tilde N}$. The probability that the system is in state $N$ at time $t$ is denoted $P(N,t)$ and the transition rate for a change from state $N$ to state ${\tilde{N}}$ is $\mathcal{T}_{\tilde{N}|N}$. In what follows, we suppress the time dependence of $N_s(k,t)$ whenever it is clear that $N_s(k)$ depends on the current time $t$. We will consider two types of transition rates $\mathcal{T}_{\tilde{N}|N}$, one describing a random walk and the second local reactions:
\begin{align}\label{eq:transitionRatesGeneral}
\mathcal{T}_{N_s(k)-1,N_s(l)+1|N_{s}(k),N_s(l)} &= \frac{D_s}{h^2}N_s(k),\quad l=k\pm 1 \nonumber\\
\mathcal{T}_{N_{1}(k)+\rho_{1,r},\dots,N_{s_{max}}(k)+\rho_{s_{max},r}|N_{1}(k),\dots,N_{s_{max}}(k)} &=R_r\left(N_{1}(k),\dots,N_{s_{max}}(k)\right).
\end{align}
In \eqref{eq:transitionRatesGeneral}, $k=2,\dots,k_{max}-1$, whereas the specification of the transition rates in the boundary boxes $k=1,k_{max}$ depends on the boundary conditions.
Furthermore, $r=1,\dots,M$ denotes the index of a particular reaction, and $\rho_{s,r}$ specifies how reaction $r$ changes the number of individuals of species $s$.
Thus, a step in the random walk changes the spatial distribution of a particular species, whereas reactions act locally in a particular box $k$. 

We define shift operators
\beq
E^\pm_{k,s}f(N_{s}(k),\dots):=f(N_{s}(k)\pm 1,\dots),
\eeq
where the dots indicate that the function can depend on $N_{s'}(k')$ for $s' \neq s$, $k' \neq k$, but operator $E^\pm_{k,s}$ affects only $N_{s}(k)$. Substituting \eqref{eq:transitionRatesGeneral} in \eqref{eq:generalMasterEquation}, we can write the master equation as
\begin{align}
\label{eq:generalMasterEquationOperators}
 \frac{dP}{dt} &= \sum_{i,j=\nene{i}}\left(E_{i,s}^+E_{j,s}^--1\right)\mathcal{T}_{N_{s}(i)-1,N_s(j)+1|N_{s}(i),N_s(j)}P\nonumber\\
 &+\sum_{i,r}\left(\prod_s E^{-\rho_{s,r}}_{i,s}-1\right)\mathcal{T}_{N_1(i)+\rho_{1,r},\dots,N_{s_{max}}(i)+\rho_{s_{max},r}|N_{1}(i),\dots,N_{s_{max}}(i)}P.
\end{align}
%

\subsection{Mean-Field Limit}

The mean quantities $\mean{N_s(k)}=\sum_{N_s(k)} N_s(k)P(N)$ evolve in time according
\begin{align}
 \frac{d\mean{N_s(k)}}{dt} &= \sum_{i,j\in\nene{i}}N_s(k)\left(E^+_{i,s}E^-_{j,s}-1\right)\mathcal{T}_{N_{s}(i)-1,N_s(j)+1|N_{s}(i),N_s(j)}P\nonumber\\
 &+\sum_{i,r}N_s(k)\left(\prod_s E^{-\rho_{s,r}}_{i,s}-1\right)\mathcal{T}_{N_1(i)+\rho_{1,r},\dots,N_{s_{max}}(i)+\rho_{s_{max},r}|N_{1}(i),\dots,N_{s_{max}}(i)}P\nonumber\\
 &= \frac{D_s}{h^2}\left(\mean{N_s({k+1})}+\mean{N_s({k-1})}-2\mean{N_s({k})}\right) \nonumber\\
 &+ \sum_r\rho_{s,r}\mean{R_r\left(N_{1}(k),\dots,N_{s_{max}}(k)\right)},\nonumber\\
 k&=2,\dots,k_{max}-1.
\end{align}
The equations for $k=1,k_{max}$ depend on the boundary conditions. If the reaction terms are non-linear, then these equations are not closed but depend on higher moments of $N_s(k)$. However, if the number of particles of each species is large, $N_s(k)\propto \Omega$, where $\Omega\gg1$, we can perform a system size expansion in inverse powers of $\Omega$ \cite{van1992stochastic}. The leading order term in the expansion closes the time evolution equations for the means so that
\begin{align}
\label{eq:generalMeanField}
 \frac{d\mean{N_s(k)}}{dt} &= \frac{D_s}{h^2}\left(\mean{N_s({k+1})}+\mean{N_s({k-1})}-2\mean{N_s({k})}\right) + \tilde{R_s}\left(\mean{N_1(k)},\dots,\mean{N_{s_{max}}(k)}\right),\nonumber\\
 k&=2,\dots,k_{max}-1.
\end{align}
By expanding $\sum_r\rho_{s,r}\mean{R_r\left((N_{1}(k),\dots,N_{s_{max}}(k)\right)}$ to lowest order in $\frac{1}{\Omega}$, we obtain the total reaction rate $\tilde{R_s}$ for species $s$. In the limit $h\to 0$ we can obtain from \eqref{eq:generalMeanField} continuum reaction-diffusion equations for the particle densities $n_s(x,t)$:
\begin{align}\label{eq:generalPDE}
 \frac{\partial n_s(x,t)}{\partial t} &= D_s\nabla^2n_s(x,t) + r_s(n_1,\dots,n_{s_{max}}).
\end{align}
In \eqref{eq:generalPDE} we have abused notation, identified $n_s(x,t) \equiv n_s(k,t) = \frac{\mean{N_s(k,t)}}{h}$ for $kh-\frac{h}{2}<x\leq kh+\frac{h}{2}$, and introduced a local reaction term $r_s$, which is obtained from $\tilde{R_s}$ by 
\beq
\tilde{R_s}\left(\mean{N^1(k)},\dots,\mean{N^{s_{max}}(k)}\right)=hr_s\left(\frac{\mean{N^1}(k)}{h},\dots,\frac{\mean{N^{s_{max}}}(k)}{h}\right).
\eeq
Note that in higher spatial dimensions ($d\geq 2$), the $h$ in the denominator needs to be replaced by $\frac{1}{h^d}$. Furthermore, we note that to solve \eqref{eq:generalPDE} numerically, we must discretise the PDE, potentially with a different discretization than that used for the stochastic compartment model. This might be desirable as for a PDE, typically we want the lattice to be as small as computationally reasonable to avoid discretization errors; by contrast for the stochastic model we must be careful when simulating nonlinear reactions if the lattice becomes similar in size to the reaction radii. Hence, if we do not alter the compartmental model, we cannot decrease the compartment size arbitrarily, see also the discussion in \cite{isaacson2013convergent}. For the remainder of this paper we use the mean field equation \eqref{eq:generalMeanField}, and do not consider the continuum limit \eqref{eq:generalPDE} any further. We emphasise wheather a variable is part of the deterministic or stochastic regime by using capital letters $N_k$ for stochastic variables, and lower case letters $n_k$ for deterministic variables, so \eqref{eq:generalMeanField} is rewritten in terms of the $n_k$ as
\begin{align}
\label{eq:finiteDifferenceGeneralEuler}
 \frac{d\mean{n_s(k)}}{dt} &= \frac{D_s}{h^2}\left(\mean{n_s({k+1})}+\mean{n_s({k-1})}-2\mean{n_s({k})}\right) + r_s\left(\mean{n_1(k)},\dots,\mean{n_{s_{max}}(k)}\right),\nonumber\\
 k&=2,\dots,k_{max}-1.
\end{align}
%

\subsection{Method of Solution of the Stochastic Model}

We solve the stochastic model using the Gillespie algorithm \cite{gillespie1976general,gillespie1977exact}, exploiting the fact that the time to the next event is distributed exponentially. Hence, if $a_k$ denotes one of the non-zero transition rates of the model, and $r_1$ is a random number uniformly distributed in $[0,1]$, then
\begin{align}\label{eq:timeToNextEvent}
\tau = -\frac{1}{\sum_ka_k}\log(r_1)
\end{align}
gives the time to the next event. Furthermore, if $r_2$ is a second, independent random number uniformly distributed in $[0,1]$, we can calculate which event $l$ happens by imposing the condition
\begin{align}\label{eq:whichStochasticEvent}
\frac{\sum_{k=1}^{l-1}a_k}{\sum_{k}a_k}<r_2<\frac{\sum_{k=1}^la_k}{\sum_{k}a_k}.
\end{align}
The advantage of using the Gillespie algorithm is that it simulates exactly the defined Markov process. The hybrid methods discussed in this paper do not depend on the details of how the stochastic process is simulated, and faster algorithms such as the $\tau$ leaping algorithm \cite{gillespie2001approximate,cao2006efficient} can also be used. Since these algorithms are not exact, in this paper we prefer to keep the stochastic simulation as detailed and accurate as possible and use the original Gillespie algorithm, and improve computational performance by switching to a mean field description when appropriate.

\section{General Algorithm for Stochastic/Deterministic Reaction-Diffusion Hybrid Model}\label{sec:generalAlgorithm}

Our algorithm involves decomposing the spatial domain into two regions for each species. In one region the species of interest is modelled in a stochastic way; in the other the mean-field limit is used. In regions where one species is modelled stochastically and the other deterministically, all interaction terms involving the two species are modelled stochastically. The interface condition describes how the two domains are coupled together, and how the interface moves.
The following steps are performed by the hybrid model repeatedly:
\\

{\bf Hybrid Algorithm}
\begin{enumerate}
 \item Generate time to next stochastic event via equation \eqref{eq:timeToNextEvent}
 \item Simulate which stochastic event happens via condition \eqref{eq:whichStochasticEvent}
 \item Iterate finite difference scheme to new time via equation \eqref{eq:finiteDifferenceGeneralEuler}
 \item Calculate interface condition
 \item Return to Step 1 until specified end time
\end{enumerate}

For notational simplicity, in the main part of this section, we focus on a one-species system with a single interface. We explain in \ref{app:alterationsAlgorithm} how this algorithm can be modified to account for multiple species and multiple dynamic interfaces.

\subsection{Hybrid Algorithm with a Single Interface}\label{subsec:singleParticlesingleInterface}

Let $[0,L]$ be our modelling domain, discretized as before into $k_{max}$ compartments of size $h$ such that $k=1,\dots,k_I-1$ is the mean field domain, $k=k_I+1,\dots,k_{max}$ is the stochastic domain, and $k=k_I$ labels the interface compartment. We use equation \eqref{eq:finiteDifferenceGeneralEuler} to solve for the variables $n_s(1),\dots,n_s({k_{I-1}})$ in the deterministic regime, whereas the evolution of the stochastic variables $N_s({k_{I+1}}),\dots,N_s({k_{max}})$ is determined by simulations of the master equation \eqref{eq:generalMasterEquation} with transition rates \eqref{eq:transitionRatesGeneral}. The equations used to determine the variables at $k_I$ will be discussed below.
\begin{figure}[H]
\centering
\includegraphics[scale=0.4]{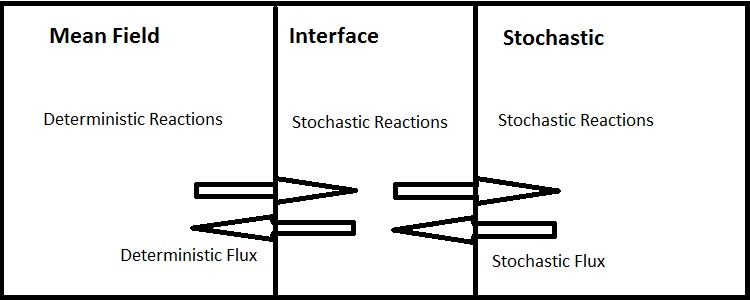}
\caption{\label{fig:domainSchematics} The domain is decomposed into a deterministic region where the system is described by the mean field equations, and a stochastic region in which it is described by the stochastic equations. At the interface, which is a single compartment, between these domains the flux into the mean field domain is deterministic (equation \eqref{eq:finiteDifferenceInterfaceFlux}), whereas the reactions and flux into the stochastic domain are calculated in a stochastic way (equations \eqref{eq:transitionRatesInterface} and \eqref{eq:transitionRatesInterfaceReaction}).}
\end{figure}
%

\subsubsection{Fluxes and Reactions at the Interface}\label{sec:interfaceConditionGeneral}

We now explain how the stochastic and deterministic regimes are coupled at the interface. We identify at all times
\begin{align}
 n_s({k_{I}})=\frac{N_s({k_I})}{h},
\end{align}
to emphasise that the interface compartment will exhibit both deterministic and stochastic behaviour.
Three processes contribute to changes in particle numbers in compartment $k_I$: fluxes into and from compartment $(k_I-1)$, which is part of the mean field domain, fluxes into and from compartment $(k_I+1)$, which is part of the stochastic domain, and local reactions (see Figure \ref{fig:domainSchematics}). Hence, we model the flux between compartments $k_I$ and $(k_I-1)$ deterministically, and the flux between compartments $k_I$ and $(k_I+1)$ in a stochastic manner. If $\tau$ denotes the current Gillespie time step, we calculate
\beq
\label{eq:finiteDifferenceInterfaceFlux}
n_s({k_{I}},t+\tau) = n_s(k_{I},t) + \tau \frac{D_s}{h^2}\left(n_s(k_I,t)-n_s(k_{I}-1,t)\right).
\eeq
The flux between boxes $k_I$ and $(k_{I}+1)$ is accounted for by the transition rates 
\begin{align}
\label{eq:transitionRatesInterface}
\mathcal{T}_{N_s(k_I)-1,N_s(k_I+1)+1|N(k_I),N(k_I+1)} &= \frac{D}{h^2}N_s(k_I),\nonumber\\
\mathcal{T}_{N_s(k_I)+1,N_s(k_{I}+1)-1|N_s(k_I),N_s(k_I+1)} &= \frac{D}{h^2}N_s(k_I+1).
\end{align}
We also have
\begin{align}
\mathcal{T}_{N_s(k_I)\pm 1,N_s(k_{I}-1)\mp 1|N_s(k_I),N_s(k_I-1)} &= 0,
\end{align}
as the corresponding flux is already accounted for by equation \eqref{eq:finiteDifferenceInterfaceFlux}.
Finally, we specify the local reactions in a stochastic way via transition rates 
\beq
\label{eq:transitionRatesInterfaceReaction}
\mathcal{T}_{N_{1}(k_I)+\rho_{1,r},\dots,N_{s_{max}}(k_I)+\rho_{s_{max},r}|N_{1}(k_I),\dots,N_{s_{max}}(k_I)} = R^r(N_{1}(k_I),\dots,N_{s_{max}}(k_I)).
\eeq
In \ref{app:deterministicInterfaceReaction} we discuss an alternative formulation for which local reactions are calculated in the mean field framework.

We remark that due to the interfacial coupling, the mean field solution acquires some stochasticity. Indeed, formally, \eqref{eq:finiteDifferenceInterfaceFlux} appears to correspond to a Neumann-like boundary condition at the interface. However, since $n_s(k_I)$ is subject to both stochastic reactions and a stochastic flux into the stochastic domain, the interface condition appears as a stochastic source for the mean field model at the interface. 

The mean behaviour of the full hybrid model is obtained by solving the mean field equations in the whole domain. By construction, these are obtained from the mean of the stochastic model. We now have to convince ourselves that the mean behaviour at the interface gives the mean field limit of the full stochastic model. Note that \eqref{eq:finiteDifferenceInterfaceFlux} appears to diverge in the limit $h\to 0$. This is because the deterministic contribution due to diffusion includes only the flux between the interface and the deterministic domain. To obtain the full discrete Laplacian, we must add the mean flux to the right-hand side. This is easily obtained if we calculate the mean field limit associated with the transition rates \eqref{eq:transitionRatesInterface}, giving $\frac{D_s}{h^2}\left(n_s(k_I+1,t)-n_s(k_{I},t)\right)$, and this contribution is simply added to \eqref{eq:finiteDifferenceInterfaceFlux}.

\paragraph{Fractional particles}

Conventionally, the state of the stochastic model is defined by the numbers of individuals $N_s(k)$ in each compartment $k$, and these are non-negative integers, whereas the densities $n_s(k)$ are real valued. At the interface the flux into and from the mean field domain is given by equation \eqref{eq:finiteDifferenceInterfaceFlux} which alters $n_s({k_I})$, and thus $N_s({k_I})$, by a real valued number. However, local reactions, as well as the flux into and from the stochastic domain, are described by transition rates \eqref{eq:transitionRatesInterface} and \eqref{eq:transitionRatesInterfaceReaction}, which effect integer changes in $N_s(k_I)$. Consequently, it is not a priori clear whether the stochastic components of the reactions and fluxes at the interface are well defined. First, we note that, by definition, the interface is such that the particle number there is sufficiently large, $N_s(k_I)\gg 1$, so that the mean field description is accurate, and hence agrees closely with the corresponding stochastic model (which has integer valued particle numbers). Adding a real part between zero and one to a large integer number will not significantly alter the transition rates, so the stochastic model with real valued $N_s(k_I)$ will agree closely with both the stochastic model with integer valued $N_s(k_I)$ and the mean field model. Formally, the state space of the stochastic model at the interface is thus still in one-to-one correspondence with the integers, which are shifted by the fractional part of $N_s(k)$. Hence, the stochastic part of the hybrid model is well-defined. There is still a numerical mismatch between the results obtained from a Gillespie algorithm with fractional or with integer valued numbers, but as long as $N_s(k_I)\gg 1$, we found this mismatch to be negligible.

\subsection{Moving Interface Condition}

The condition used to locate the interface is not dictated by a rigorous mathematical requirement: it represents a compromise between performance and accuracy. We view the mean field equations as an approximation to the stochastic model that neglects fluctuations. Hence, the larger the mean field domain, the fewer stochastic fluctuations are taken into account. However, we then typically increase the performance of the simulation. 

We determine whether a compartment belongs to the stochastic or mean field domain by comparing the number of particles in that compartment with a threshold number, $\Theta_s$, such that if $N_s(k) < \Theta_s$, then box $k$ is part of the stochastic domain, and otherwise part of the mean field domain\footnote{If this condition splits the whole modelling domain such that the stochastic domain consists of multiple, disconnected components we either need to introduce multiple interfaces, see \ref{app:multipleInterfaces}, or connect the disconnected components.}. This is justified since fluctuations typically scale with the square root of the number of particles in a box, $\sqrt{N_s(k)}$. As the number of individuals may change over in time, the position of the interface may also evolve in space and time.

We also implement a minimum domain size condition as a simple check that no connected component of the mean field domain is allowed to become too small. Imagine, for instance, that the mean field domain is enclosed by two disconnected components of the stochastic domain. If the mean field domain comprises only a few compartments in the discretisation, it might be computationally more efficient to remove the mean field domain and absorb it into the stochastic domain, as then we do not have to calculate the interface condition. In the simulations performed in this paper a minimum domain size of $5$ compartments for the mean field model was used. In higher dimensions, we anticipate that a cube with a length of $5$ compartments would work well. We stress, however, that the choice of threshold conditions is model specific, and hence the minimal domain size requirement should be chosen on a case-by-case basis.

After the position of the interface is updated, we must check that all particle numbers in the stochastic domain are integer values, paying particular attention to compartments that were previously part of the mean field domain. By mass conservation, this will result in a renormalisation of the density functions in the mean field domain: this procedure is discussed in section \ref{sec:Renormalisation}.

The following steps are used to adjust the position of the interface:
\begin{enumerate}
 \item Calculate threshold condition to locate interface;
 \item Calculate minimum domain size condition;
 \item Renormalise particles and densities.
\end{enumerate}
%

\subsection{Renormalision of Particle Distribution}\label{sec:Renormalisation}

When the interface moves such that a compartment previously treated deterministically and, hence, described by real valued densities $n_s(k)$, enters the stochastic domain, we need to ensure that the number of particles becomes integer valued. By mass conservation, we cannot simply remove the fractional part of $N_s(k)=hn_s(k)$: instead we rescale the densities outside the stochastic domain.

Let us focus on a single box $k$ (and assume that the single interface has moved by only one compartment, so that $N_s(k)$ is now non-integer valued but part of the stochastic domain). We interpret the fractional part
\beq
 p_s = N_s(k) \mod(1),
\eeq
as the probability that an additional particle is in this box. We draw a uniform random number $r\in[0,1]$. If $r<p_s$ then we place the particle in box $k$; otherwise it is placed in the deterministic domain. To preserve particle numbers, we renormalise the density function in the deterministic regime and reset $N_s(k)$ so that if $r<p_s$ then
\begin{align}\label{eq:renormalisation1}
 N_s(k)\ &\to N_s(k)+(1-p_s),\nonumber\\
 n_s(l) &\to \left(1-\frac{(1-p_s)}{\sum_{m=1}^{k-1}n_{s}(m) h}\right) n_s(l), \quad l=1,\dots,k-1,
\end{align}
and otherwise
\begin{align}\label{eq:renormalisation2}
 N_s(k) &\to N_s(k)-p_s,\nonumber\\
 n_s(l) &\to \left(1+\frac{p_s}{\sum_{m=1}^{k-1}n_s(m) h}\right) n_s(l), \quad l=1,\dots,k-1.
\end{align}
A similar rescaling procedure was used in \cite{franz2013}, where a reaction-diffusion PDE was coupled with a Brownian dynamics model. There, it was found that particles crossing the interface twice can cause increased variance. Our algorithm does not lead to an observable increase in variance as, by construction, the interface is chosen so that the mean field and interface domain always contain a large number of particles.

\section{Stochastic Fisher-Kolmogorov Equation}\label{sec:Fisher}

In this section, we apply our algorithm to the Fisher-Kolmogorov equation,
\begin{align}\label{eq:Fishereqn}
\frac{\partial n}{\partial t} = D\frac{\partial^2 n}{\partial x^2} +\lambda n(1-\frac{n}{\Omega})
\end{align}
Here, $D$ is the diffusion coefficient, $\lambda$ the growth rate and $\Omega$ the carrying capacity.

A stochastic, lattice-based version of equation \eqref{eq:Fishereqn} was studied in \cite{breuer1994fluctuation,breuer1995macroscopic} and is defined by a master equation with the transition rates
\begin{align}\label{eq:transitionRatesFisher}
\mathcal{T}_{N(k)-1,N({k\pm 1})+1|N(k),N({k\pm 1})} &= \frac{D}{h^2}N(k),\nonumber\\
\mathcal{T}_{N(k)+1|N(k)} &= \lambda N(k),\nonumber\\
\mathcal{T}_{N(k)-1|N(k)} &= \frac{\lambda}{ \Omega h} N(k) (N(k)-1), \quad k=1,\dots,k_{max},
\end{align}
with Dirichlet boundary conditions $N_1=\Omega$, $N_{k_{max}}=0$.
The evolution of the means $\mean{N(k)}$ is given by
\begin{align}\label{eq:meanEquationFisher}
\frac{\partial \mean{N(k)}}{\partial t} = \frac{D}{h^2}\left(\mean{N({k+1})}-2\mean{N(k)}+\mean{N({k-1})}\right) + \lambda\mean{N(k)\left(1-\frac{N(k)}{\Omega h}\right)}.
\end{align}
Defining $n(x,t)=\frac{\mean{N(k,t)}}{h}$, with $x=hk$, the mean field equation for $n$ reduces to \eqref{eq:Fishereqn} if $\Omega\gg 1$, as the van Kampen approximation at leading order implies the moment reduction $\mean{N(k,t)^2}\approx\mean{N(k,t)}^2$. 

We will now use the hybrid algorithm to simulate travelling wave solutions as we vary several model parameters. As noted in \cite{breuer1994fluctuation,breuer1995macroscopic} for the stochastic Fisher-Kolmogorov equation with the conventions used in the present paper and in \cite{moro2004,brunet2001effect,conlon2005travelling} for alternative formulations, stochastic effects can produce wave speeds $c_{stoch}$ which deviate from the deterministic Fisher-Kolmogorov equation \eqref{eq:Fishereqn}, $c_{PDE}=2\sqrt{D\lambda}$. The three other parameters control the various limits: $\Theta\to\infty$ yields the stochastic model from the hybrid model, $\Omega\to\infty$ yields the mean field model from the stochastic model, and $h\to0$ yields the PDE from the mean field model. We will study the effect of variation in these three parameters on travelling wave speeds in the next subsection.

\subsection{Fisher-Kolmogorov Travelling Waves}

We now study travelling wave solutions, fixing $D=1$, $\lambda=1$. We ensure that in all simulations the travelling wave is sufficiently far from the boundaries, so that effects associated with the finite domain size are negligible. As initial conditions we approximate the travelling wave solution of the PDE (\ref{eq:Fishereqn})), so that we can focus on wave propagation, rather than wave formation.
\begin{figure}[h!]
\subfloat[Stochastic model, $\Omega=10$]
{
\label{fig:TravellingWaveStochastic_10}
\includegraphics[width=0.48\linewidth]{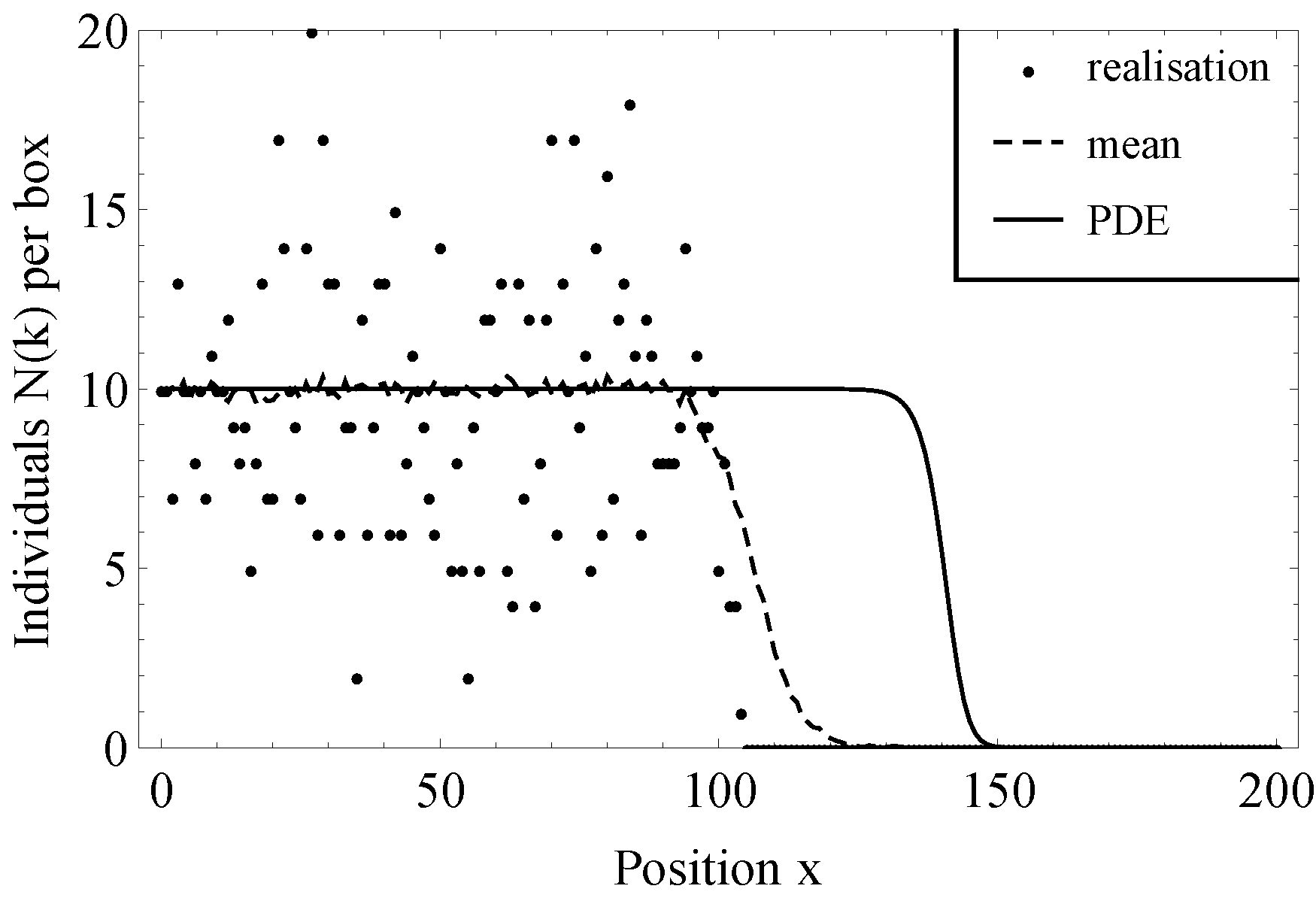}
}
\subfloat[Hybrid model, $\Theta=25$, $\Omega=10$]
{
\label{fig:TravellingWaveHybrid25_10}
\includegraphics[width=0.48\linewidth]{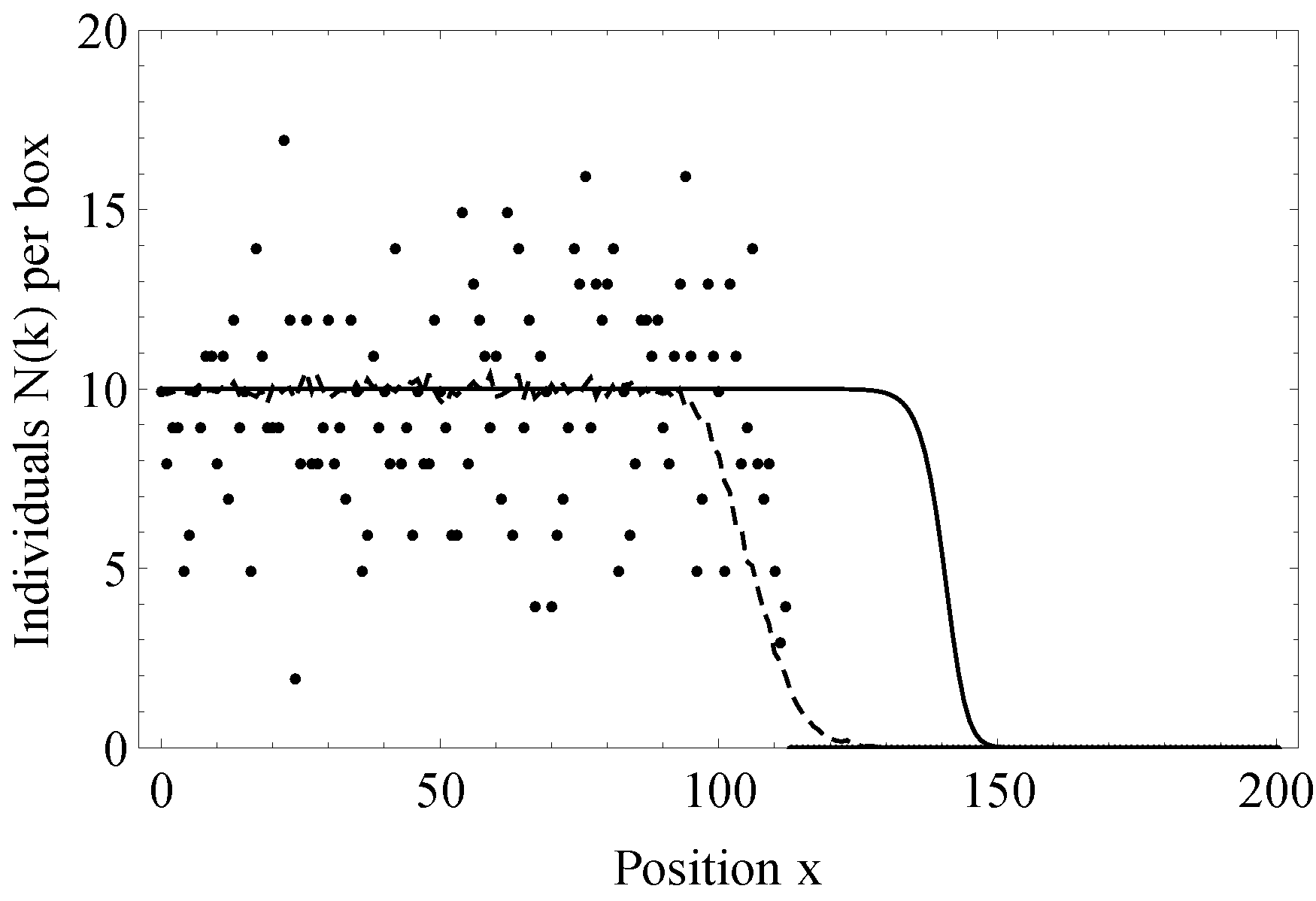}
}\\
\subfloat[Stochastic model, $\Omega=25$]
{
\label{fig:TravellingWaveStochastic_25}
\includegraphics[width=0.48\linewidth]{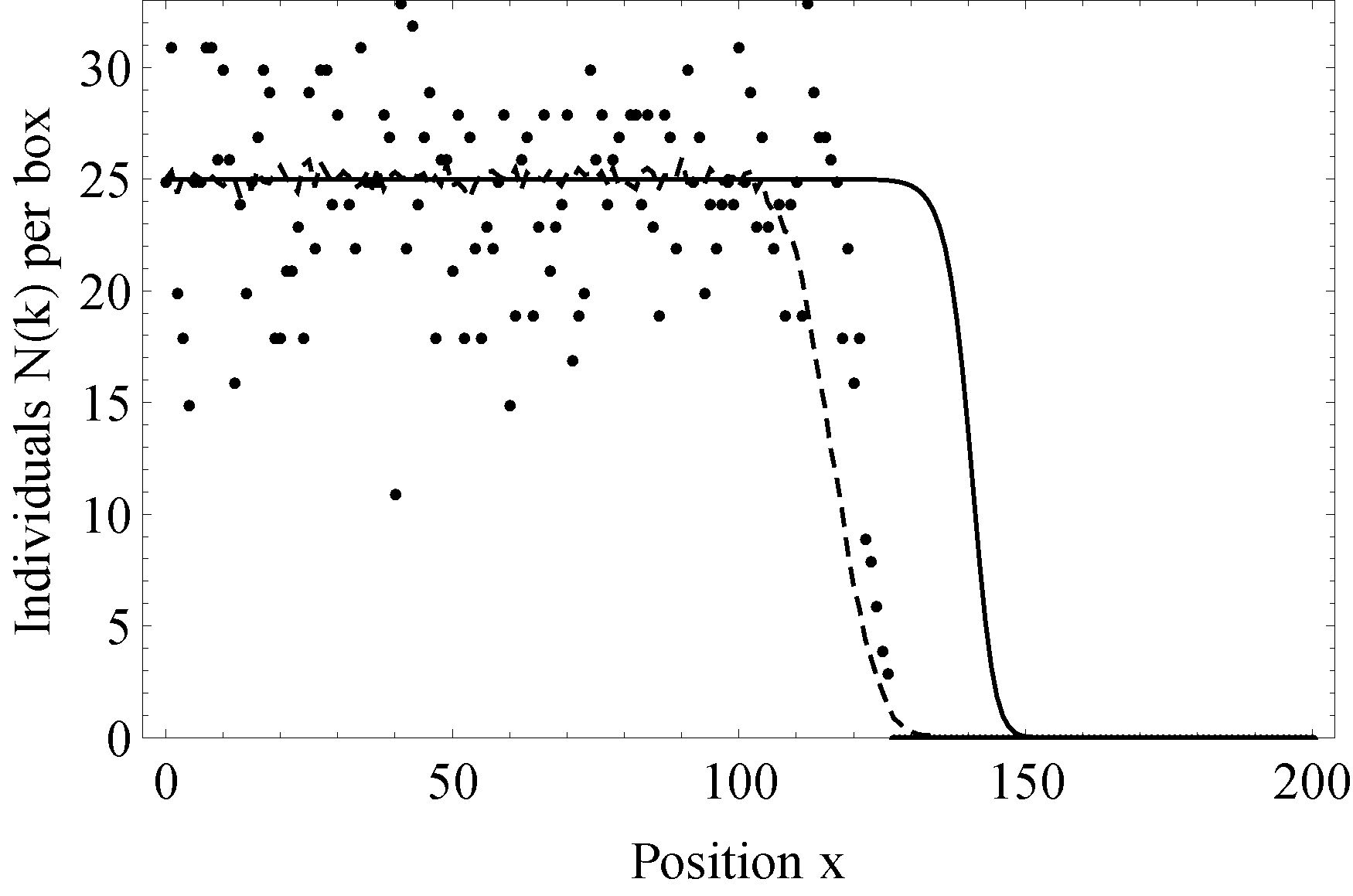}
}
\subfloat[Hybrid model, $\Theta=25$, $\Omega=25$]
{
\label{fig:TravellingWaveHybrid25_25}
\includegraphics[width=0.48\linewidth]{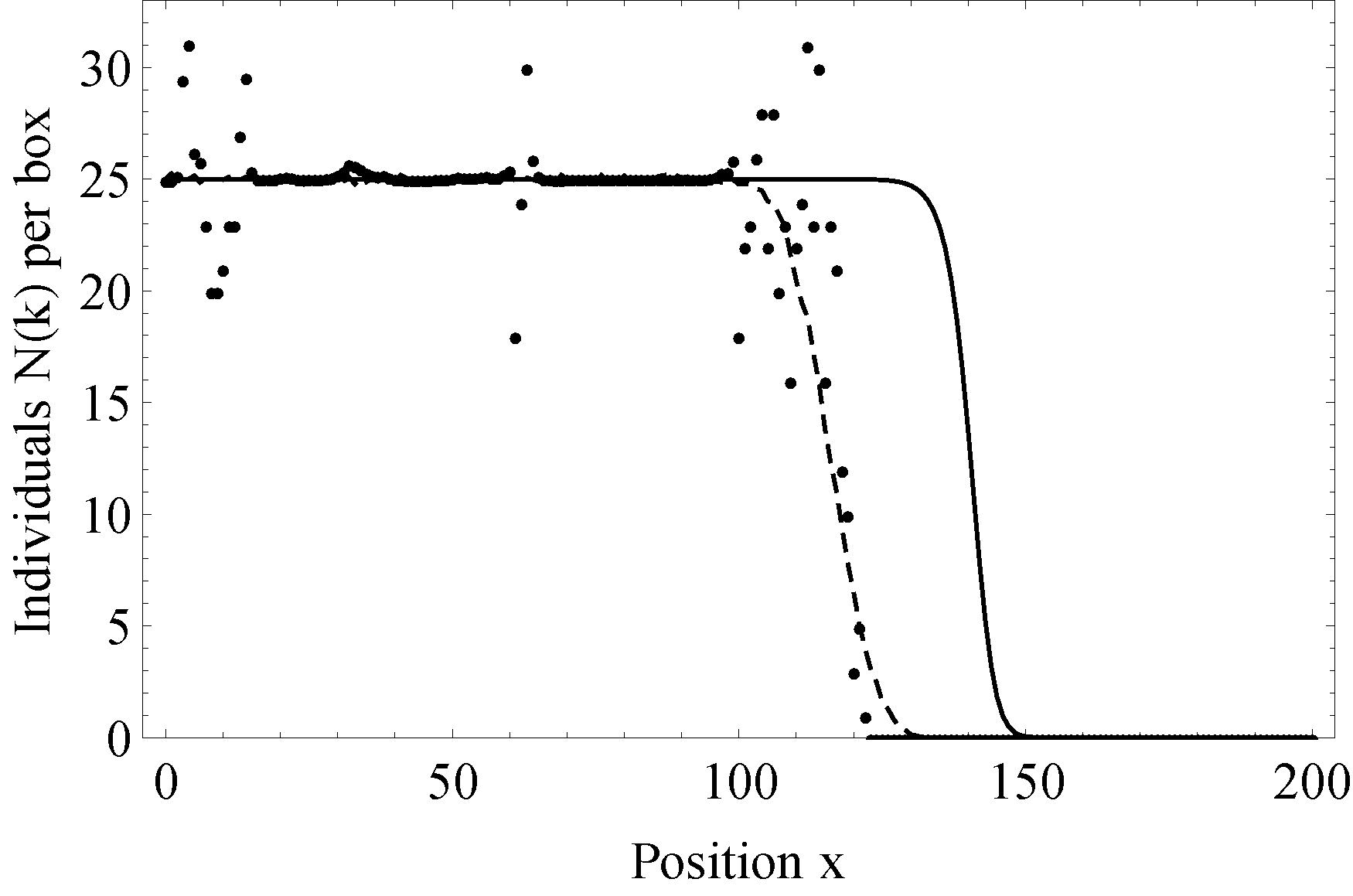}
}\\
\subfloat[Stochastic model, $\Omega=50$]
{
\label{fig:TravellingWaveStochastic_50}
\includegraphics[width=0.48\linewidth]{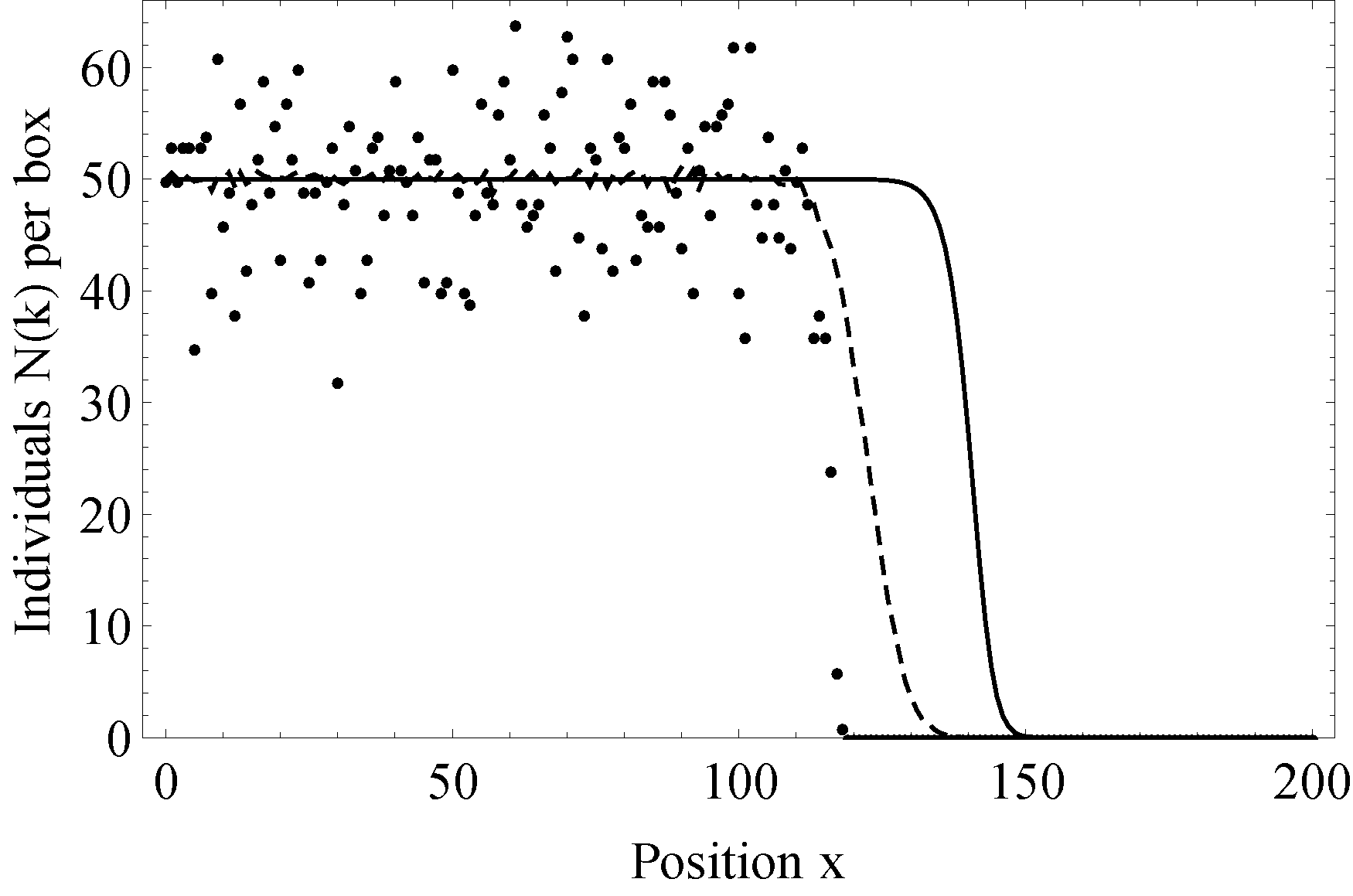}
}
\subfloat[Hybrid model, $\Theta=25$, $\Omega=50$]
{
\label{fig:TravellingWaveHybrid25_50}
\includegraphics[width=0.48\linewidth]{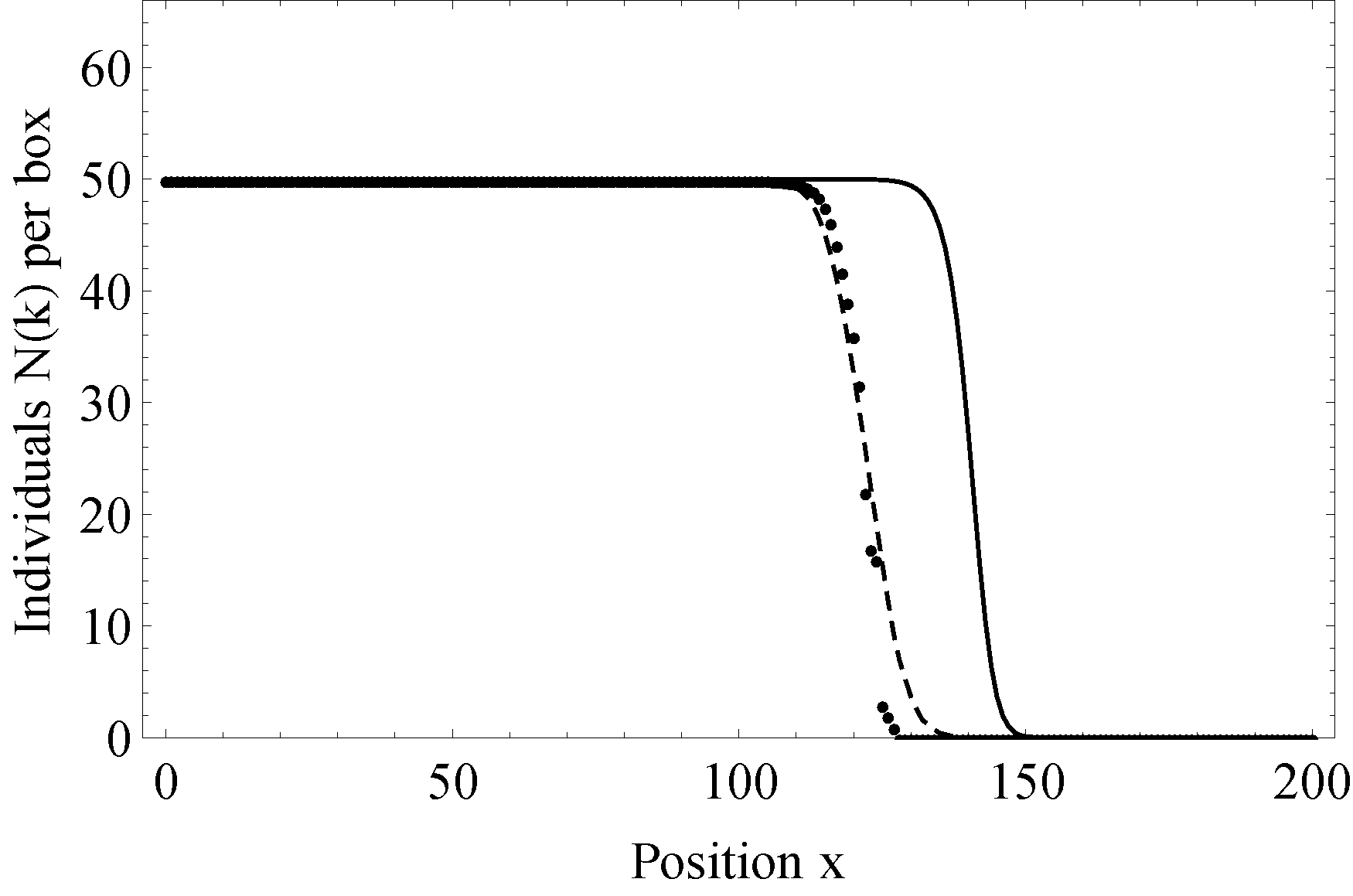}
}
\caption{\label{fig:TravellingWaves} Series of plots comparing the travelling waves profile generated by the stochastic model in the column to the left and the hybrid model with a threshold of $\Theta=25$ on the right (simulation time $t=60$). As the carrying capacity $\Omega$ increase, $\Omega=10,25,100$. Each plot shows a single realisation as well as the mean of $256$ realisations of the stochastic or hybrid model, respectively. Results for the corresponding PDE \eqref{eq:Fishereqn} are also shown. The other parameter values are $D=\lambda=1,k_{max}=20, h=1$.}
\end{figure}

Figure \ref{fig:TravellingWaves} compares travelling wave solutions generated from the stochastic model, the hybrid model (with $\Theta=25$) and the PDE. In each plot we present a single realisation and the mean of $256$ realisations of the stochastic (Figure \subref*{fig:TravellingWaveStochastic_10},\subref{fig:TravellingWaveStochastic_25},\subref{fig:TravellingWaveStochastic_50}) and hybrid models (Figure \subref*{fig:TravellingWaveHybrid25_10},\subref{fig:TravellingWaveHybrid25_25},\subref{fig:TravellingWaveHybrid25_50}) together with the numerical solution of the corresponding PDE. We fix $D=\lambda=1,k_{max}=20, h=1$, and allow to vary $\Omega$. We note that the travelling wave speeds for the stochastic and hybrid models are slower than those of the PDE, and the speed increases with $\Omega$. Furthermore, the relative noise, i.e. the fluctuation of a single stochastic or hybrid realisation about the mean, decreases as $\Omega$ increases. Finally, the wave front of the PDE appears to be steeper compared to the wave front of the mean of $256$ realisations both in the stochastic and hybrid models compared to the PDE, and the steepness increases with $\Omega$. This is explained as the different realizations of the stochastic model can have different speeds, hence the average broadens the wave front.

We now compare the stochastic to the hybrid model. For $\Omega = 10$, so $\Omega < \Theta$, neither single realisations nor the mean of the stochastic model (Figure \subref*{fig:TravellingWaveStochastic_10}) differ significantly from the hybrid model (Figure \subref*{fig:TravellingWaveHybrid25_10}), as the threshold of the hybrid model is considerably larger than the carrying capacity, so almost certainly the entire domain of the hybrid model will be stochastic. When $\Omega=25$, so $\Omega=\Theta$, the stochastic model (Figure \subref*{fig:TravellingWaveStochastic_25}) and the hybrid model (Figure \subref*{fig:TravellingWaveHybrid25_25}) differ significantly away from the wave front. The stochastic model is much noisier, but the noise in the hybrid model is non-zero as noise from the stochastic domain can diffuse into the mean field domain, raising the particle number above the threshold value. Note that fluctuations can also reach beyond the threshold in the hybrid model due to the minimum domain size requirement for the deterministic domain. When $\Omega=50$, so $\Omega > \Theta$, the noise a way from the wave front associated with the stochastic model (Figure \subref*{fig:TravellingWaveStochastic_50}) is absent in the hybrid model (Figure \subref*{fig:TravellingWaveHybrid25_50}). Nevertheless, the wave fronts of the means appear similar. Hence, for the parameter values used, the hybrid model represents a good approximation to the stochastic model, producing travelling waves with the same speed. If the carrying capacity is larger than the threshold $\Theta$ then fluctuations around the carrying capacity behind the front are suppressed, without affecting the wave speed.
\begin{figure}[h!]
\center
\includegraphics[width=0.48\linewidth]{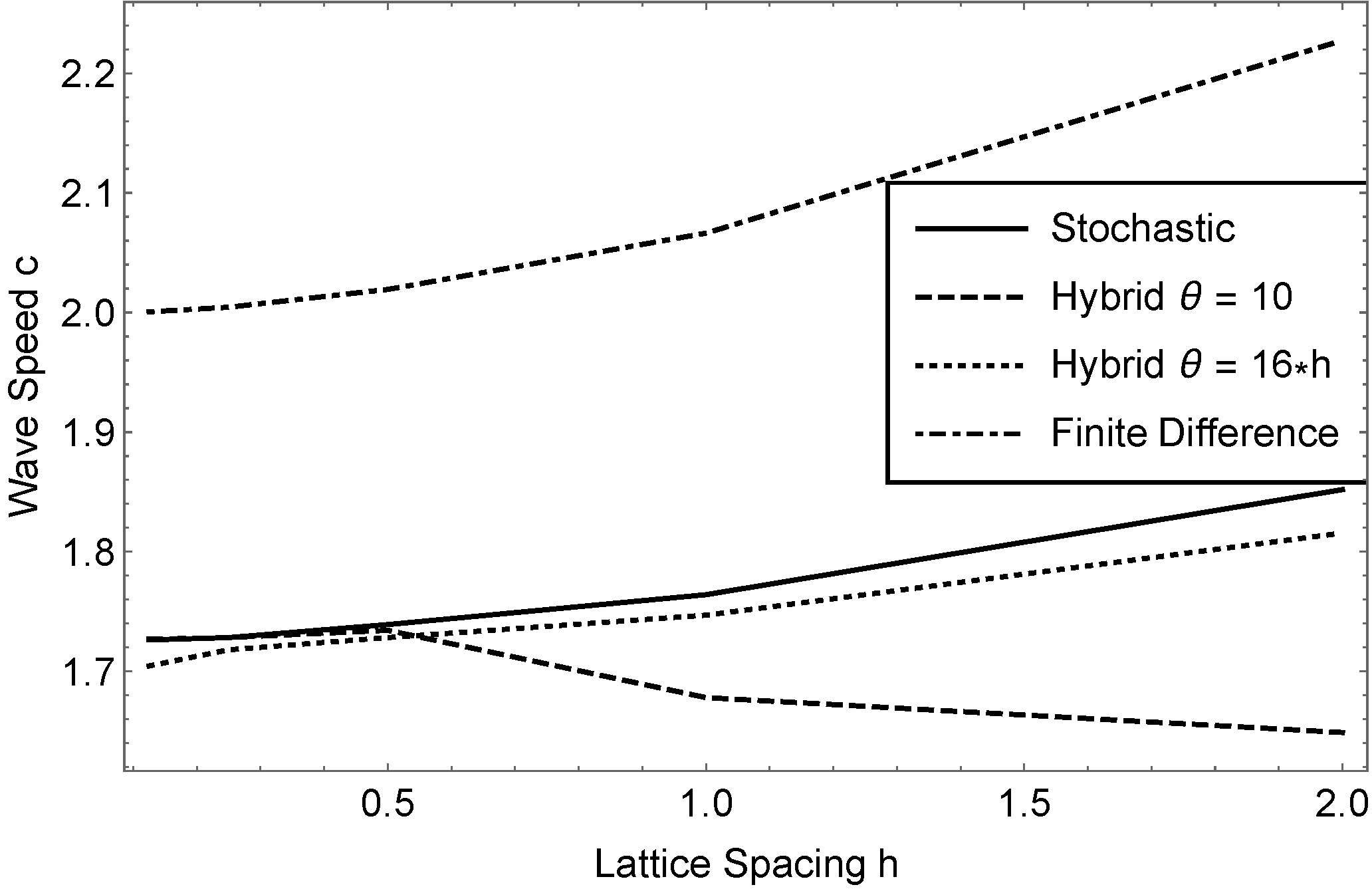}
\caption{\label{fig:TravellingWavespeedLatticeDependence}Series of curves showing how the wave speed of the stochastic model, the mean field model (i.e. the finite difference discretisation of the PDE) and hybrid models for thresholds of $\Theta=10$ and $16*h$ changes as the lattice spacing varies. We note that the wave speed of the hybrid model, with a fixed threshold, converges to that of the stochastic model, whereas the hybrid model where the threshold is adjusted with the lattice spacing does not. The other parameters are $\Omega=80*h, D=\lambda=1$, and all stochastic and hybrid results are obtained from averaging $1024$ different simulations.}
\end{figure}

Figure \ref{fig:TravellingWavespeedLatticeDependence} shows the dependence of the wave speed on the lattice constant. We compare the stochastic model against the mean field model, i.e. the finite difference discretization of the PDE with the same lattice constant, and the hybrid model. Densities are fixed by adjusting $\Omega = 80*h$. Likewise, $\Theta = 16*h$, but we also compare to the hybrid model with a fixed threshold $\Theta=10$. The wave speed $c$ is calculated by observing that the change of the total number of particles in time, averaged over all simulations, $\mean{N_{tot}}=\sum_{k=1}^{k_{max}}\mean{N(k)}$, should be proportional to $c$. We approximate the wave speed by comparing $N_{tot}$ after fixed time intervals $\Delta t=5$, obtaining $N_{tot}(t+\Delta t)-N_{tot}(t) = \frac{\Delta t \Omega}{h}c$. The finite difference model converges, as expected, to $c=2$ as $h\to0$, even though every finite lattice spacing will still result in some visible dispersion after a long time. The stochastic model is significantly lower the wavespeed while the wave speed of the hybrid model, with $\Theta=10$, converges to that for the stochastic model. However, if $\Theta=16 h$, then the hybrid model does not appear to converge to the stochastic model.
\begin{figure}[h!]
\subfloat[]
{
\label{fig:TravellingWavespeedCapacity}
\includegraphics[width=0.48\linewidth]{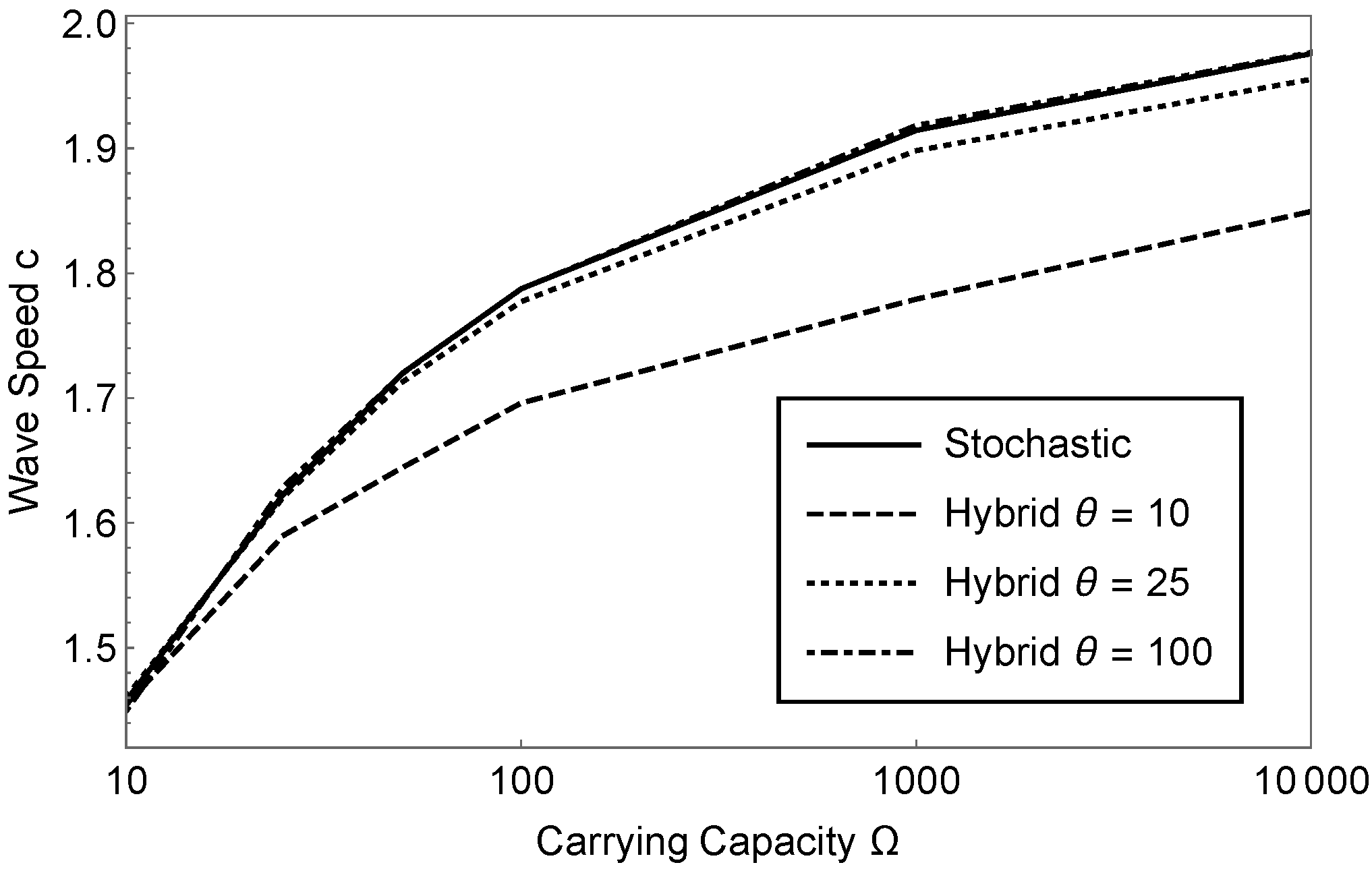}
}
\subfloat[]
{\label{fig:TravellingWavespeedThreshold}
\includegraphics[width=0.48\linewidth]{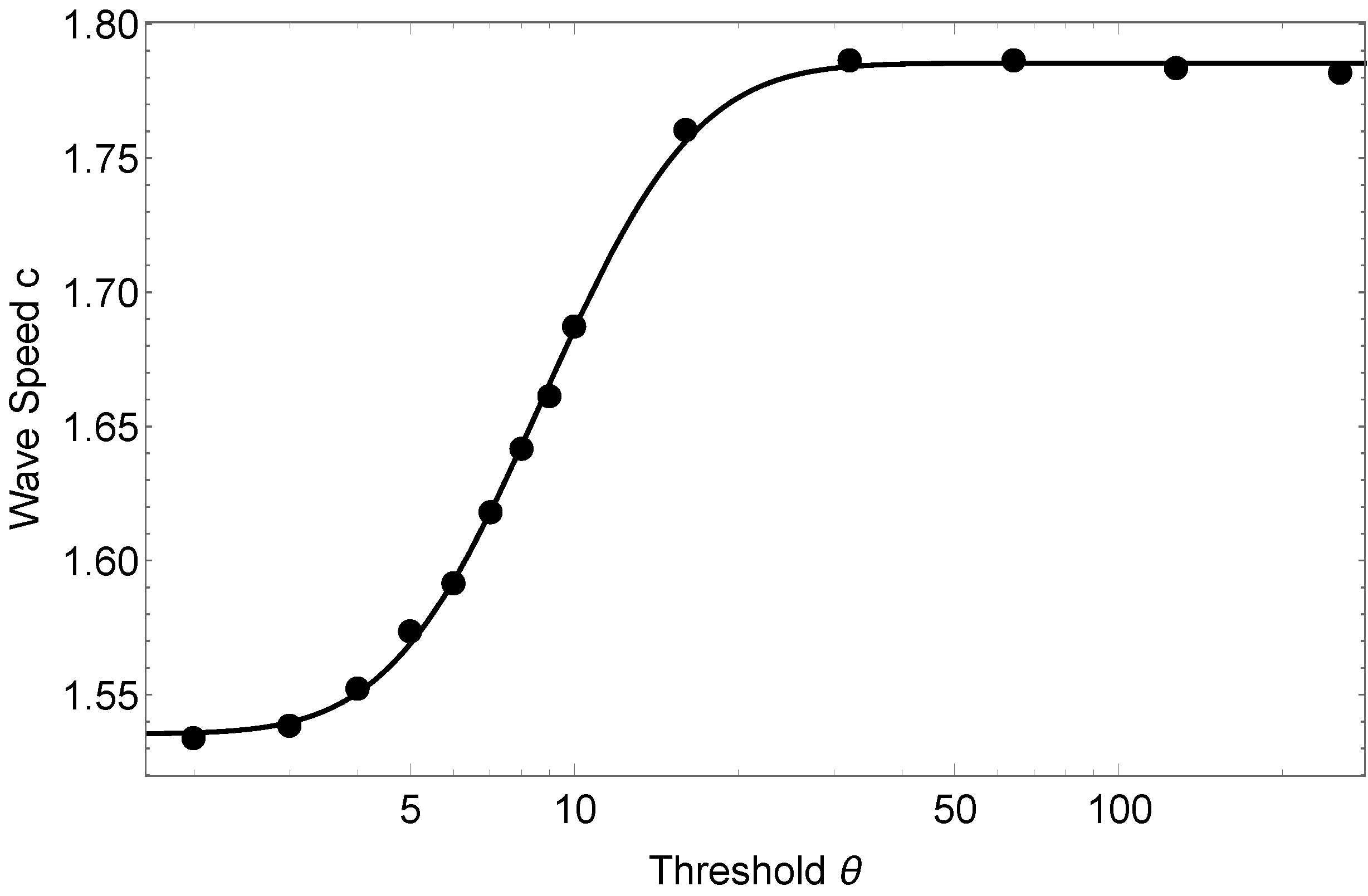}
}
\caption{\label{fig:TravellingWavespeed}The average wave speed dependence on the carrying capacity is shown for the stochastic model, as well as the hybrid model for thresholds of $\Theta=10,25$ and $100$ in \protect\subref{fig:TravellingWavespeedCapacity}. The dashed-dotted graph of the hybrid model with a threshold of $\Theta=100$ coincides with the solid line representing the stochastic model. The wave speed of the PDE is $c=2$, but the discrete mean field equations can slightly deviate from this value in an $\Omega$-independent way. \protect\subref{fig:TravellingWavespeedThreshold} shows, for a fixed carrying capacity of $\Omega=100$, explicitly the threshold dependence for several values of $\Theta$, and the solid line is a fit via the function $c = a_1 + a_2 Erf(a_3 * Log(\Theta) + a_4)$, where $Erf$ is the error function and we obtained $a_1=1.5, a_2=0.13, a_3=1.4, a_4 = -3.0$. The other parameters are $h=D=\lambda=1$, and all stochastic and hybrid results are obtained from averaging $256$ different simulations.
}
\end{figure}

Figure \subref*{fig:TravellingWavespeedCapacity} shows how the wave speed varies with the carrying capacity $\Omega$ for the stochastic and hybrid models. Here, $h=D=\lambda=1$ in all cases. The wave speeds for the stochastic model are identical to those obtained in \cite{breuer1994fluctuation}, and for large $\Omega$ they approach the wave speed of the mean field theory as expected by general theory \cite{kurtz1970solutions,van1992stochastic}, this value is slightly above $c=2$ because $h$ is finite. When $\Theta=100$, the wave speeds for the hybrid and stochastic models are indistinguishable. We remark that a naive expectation that the hybrid model should be intermediate between the PDE and the stochastic model does not imply that the wave speed of the hybrid model should be intermediate between their wave speeds. However, a correct expectation is that as $\Theta$ increases, the agreement between the hybrid and stochastic models, including their wave speeds, increases. Figure \subref*{fig:TravellingWavespeedThreshold} shows this clearly. Here, we have fixed $\Omega=100$ and varied $\Theta$. The wave speed monotonically approaches that of the stochastic model (not shown, but it is formally obtained as $\Theta \to \infty$), and for $\Theta\simeq 25$ the wave speed of the hybrid model is almost identical to that of the stochastic model. Due to the observation in those numerical simulations that the wave speed seems to plateau for high and low values of $\Theta$, we have fitted the function $c = a_1 + a_2 Erf(a_3 * Log(\Theta) + a_4)$ to the values obtained from the simulations, and obtained a very good fit for the values $a_1=1.5, a_2=0.13, a_3=1.4, a_4 = -3.0$.

\section{Spatial Stochastic Lotka-Volterra System}\label{sec:LotkaVolterra}

We now investigate a predator-prey system predator $M$ and prey $N$. In the stochastic model, each species can jump to neighbouring lattice sites, the prey reproduce at rate $a$, predators die at rate $c$ and consume prey and reproduce at rate $b$. The transition rates are given by
\begin{align}\label{eq:transitionRatesLV}
\mathcal{T}_{N(k)-1,N(k\pm 1)+1|N(k),N({k\pm 1})} &= \frac{D_N}{h^2}N(k),\nonumber\\
\mathcal{T}_{M(k)-1,M({k\pm 1})+1|M(k),M(k\pm 1)} &= \frac{D_M}{h^2}M(k),\nonumber\\
\mathcal{T}_{N(k)+1|N(k)} &= a N(k),\\
\mathcal{T}_{N(k)-1,M(k)+1|N(k),M(k)} &= b N(k) M(k),\nonumber\\
\mathcal{T}_{M(k)-1|M(k)} &= c M(k).\nonumber
\end{align}
For simplicity, we choose the interaction reaction such that each time a prey is eaten by a predator, a single new predator is born. The mean field and continuum limit corresponds to the classical spatial Lotka-Volterra equations:
\begin{align}\label{eq:LVPDE}
\frac{\partial n}{\partial t} &= D_N\frac{\partial^2 n}{\partial x^2} + a n - b n m,\nonumber\\
\frac{\partial m}{\partial t} &= D_M\frac{\partial^2 m}{\partial x^2} + b n m - c m.
\end{align}
Here, $n=n(x,t)$ and $m=m(x,t)$ are prey and predator densities related to $N(k)$ and $M(k)$ repectively in the same way as before.

As before, we use a finite difference approximation to solve equation \eqref{eq:LVPDE}, discretising in space using the same lattice as for the stochastic model. For the time integration we use the Runge-Kutta method. All plots are normalised so that the number of predators or prey in a box is shown, rather than the corresponding density.

There are now four subdomains to consider, depending on whether each of the predator and prey evolve deterministically or stochastically. For notational simplicity we identify $N(k)= h n(k)$ and $M(k)= h m(k)$ where appropriate. We will now explain explicitly which transition rates define the stochastic model, and which PDEs correspond to the mean field model solved in the respective subdomain. The interfaces between the subdomains are as given in section \ref{sec:interfaceConditionGeneral}.

\subsubsection*{Deterministic Predator and Deterministic Prey}

In this region, we solve Equations \eqref{eq:LVPDE}, and set the transition rates \eqref{eq:transitionRatesLV} equal to zero, so no stochastic reactions occur.
\subsubsection*{Deterministic Predator and Stochastic Prey}
The deterministic equation are:
\begin{align}\label{eq:LVPDEDetPredStochPrey}
\frac{\partial n}{\partial t} = 0,\quad \frac{\partial m}{\partial t} = D_M\frac{\partial^2 m}{\partial x^2}  - c m,
\end{align}
and the transition rates are the following
\begin{align}\label{eq:LVstochDetPredStochPrey}
\mathcal{T}_{N(k)-1,N({k\pm 1})+1|N(k),N({k\pm 1})} = \frac{D_N}{h^2}N(k),\mathcal{T}_{M(k)-1,M({k\pm 1})+1|M(k),M(k\pm 1)} = 0,\nonumber\\
\mathcal{T}_{N(k)+1|N(k)} = a N(k), \mathcal{T}_{N(k)-1,M(k)+1|N(k),M(k)} = b N(k) M(k), \mathcal{T}_{M(k)-1|M(k)} = 0.
\end{align}

\subsubsection*{Stochastic Predator and Deterministic Prey}
Deterministic part equations are described below,
\begin{align}
\frac{\partial n}{\partial t} = D_N\frac{\partial^2 n}{\partial x^2} + a n ,\quad \frac{\partial m}{\partial t} = 0,
\end{align}
and the transition rates are the following
\begin{align}
\mathcal{T}_{N(k)-1,N({k\pm 1})+1|N(k),N({k\pm 1})} = 0, \mathcal{T}_{M(k)-1,M({k\pm 1})+1|M(k),M{(k\pm 1)}} = \frac{D_M}{h^2}M(k),\nonumber\\
\mathcal{T}_{N(k)+1|N(k)} = 0, \mathcal{T}_{N(k)-1,M(k)+1|N(k),M(k)} = b N(k) M(k), \mathcal{T}_{M(k)-1|M(k)} = c M(k).
\end{align}

\subsubsection*{Stochastic Predator and Stochastic Prey}

Here both species are fully stochastic and we use transition rates \eqref{eq:transitionRatesLV}.\\

We will now consider two scenarios appearing in the spatial Lotka-Volterra system, and compare the hybrid model to the stochastic and deterministic models. Both scenarios correspond to solutions of the PDE which oscillate in space and time, but in one case the oscillations bring the total number of prey so close to zero that extinction is possible in the stochastic model.

\subsection{Oscillatory Behaviour without Observable Extinction}

The domain of length $L=20$ is divided into $k_{max}=101$ boxes, so $h=0.2$. Initially it contains a spatially homogeneous distribution of prey and predators so that $N(k,t=0)=50, M(k,t=0)=5$. The model parameters are fixed so that $D_N=D_M=1$, $a=2,b=0.1,c=3$. With Neumann boundary conditions, equations \eqref{eq:LVPDE} remain spatially homogeneous at all times, and both populations oscillate in time.
\begin{figure}[h!]
\centering
\includegraphics[width=0.60\linewidth]{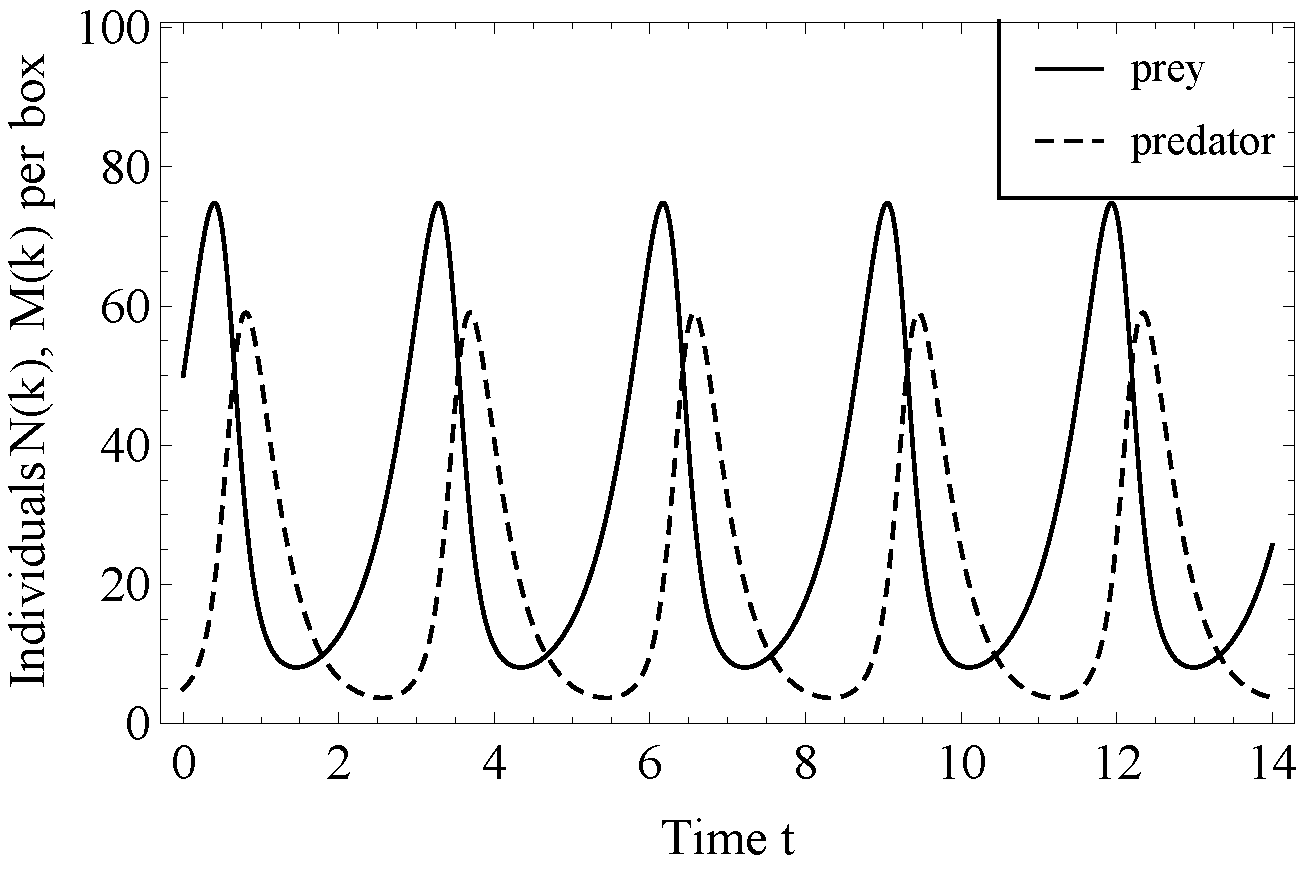}
\caption{\label{fig:flatOscillation_LVPDE}The solution of the Lotka-Volterra Eq.(\ref{eq:LVPDE}) with parameters $a=2,b=0.1,c=3$ for spatially homogeneous initial conditions $N(k,t=0)=50, M(k,t=0)=5$. Shown is the time evolution of the number of prey and predators in any given box $k$ in the discretisation, to allow for better comparison with the stochastic model. As the diffusion terms do not contribute in the spatially homogeneous case, this solution is identical to the solution of the Lotka-Volterra ODEs.}
\end{figure}
Typical results are presented in Figure \ref{fig:flatOscillation_LVPDE}, show that the peak in prey numbers is followed by a peak in the number of predators. For this choice of parameter values the minimum number of individuals of either species is always sufficiently large that extinction in the stochastic, spatial model is almost impossible.
\begin{figure}[h]
\subfloat[Spatial Profile Stochastic Model]
{
\label{fig:flatOscillationStoch_1}
\includegraphics[width=0.48\linewidth]{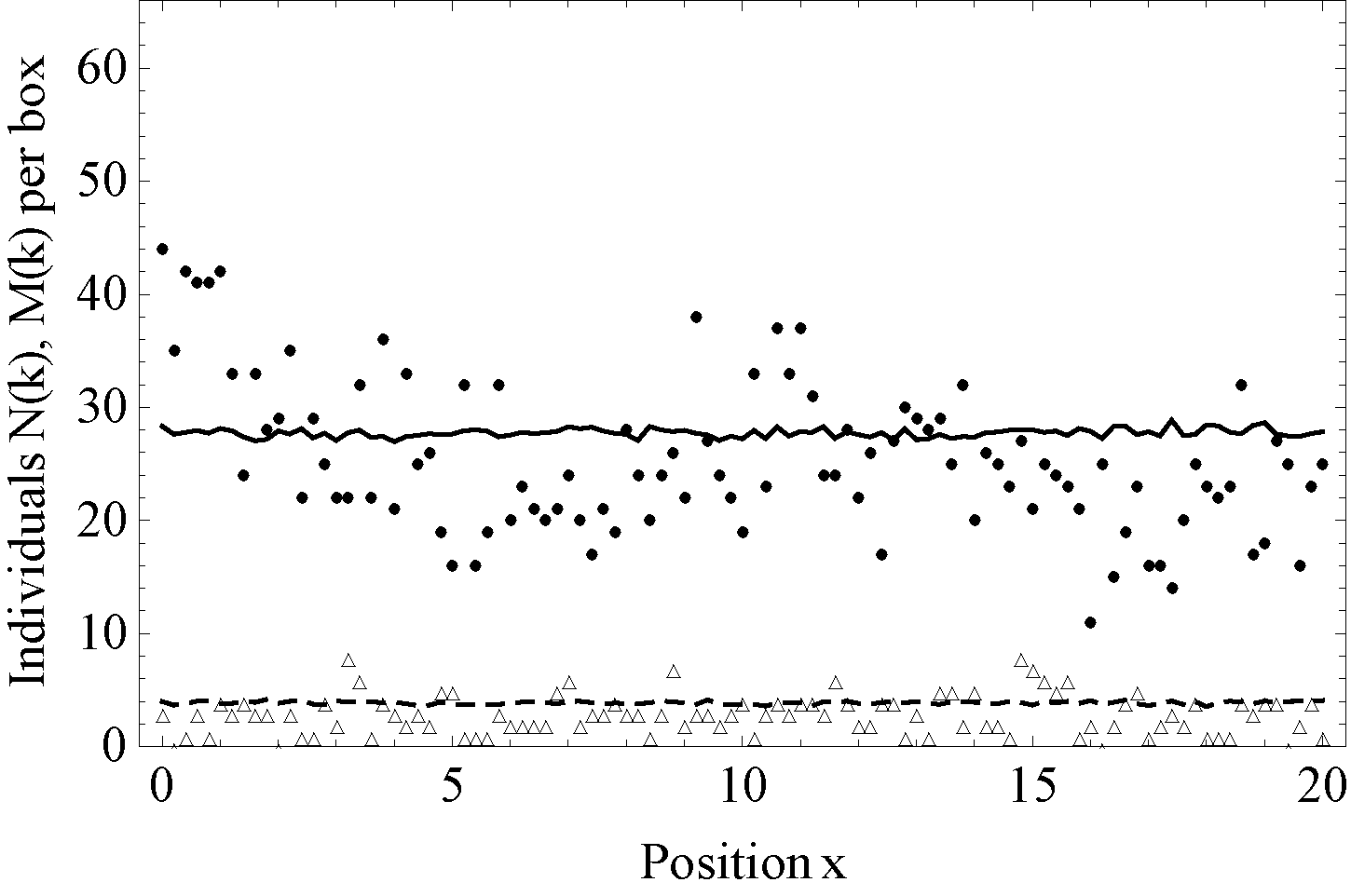}
}
\subfloat[Time Evolution Stochastic Model]
{
\label{fig:flatOscillationStoch_2}
\includegraphics[width=0.48\linewidth]{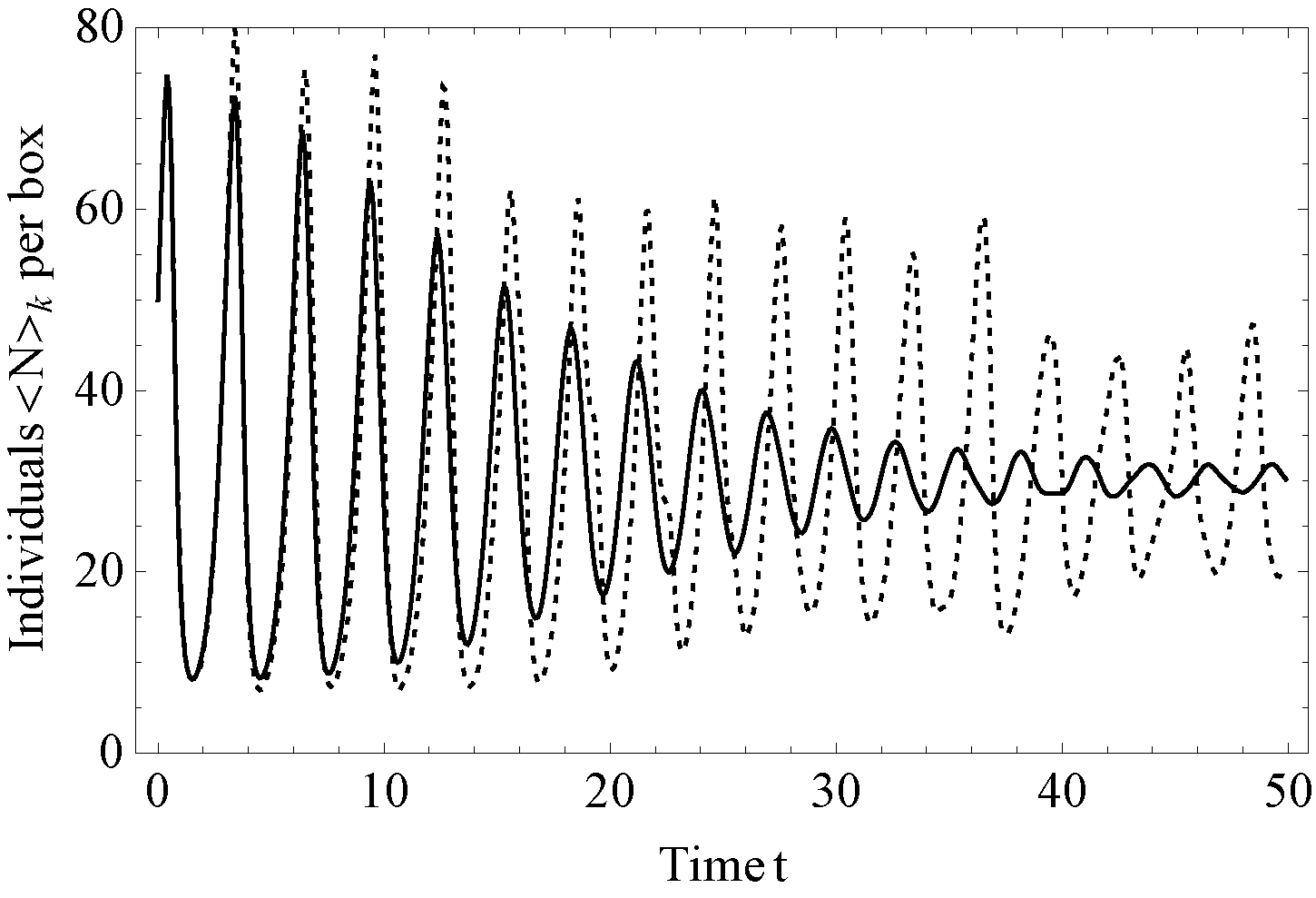}
}\\
\subfloat[Spatial Profile Hybrid $\Theta=10$]
{
\label{fig:flatOscillationHybrid10_1}
\includegraphics[width=0.48\linewidth]{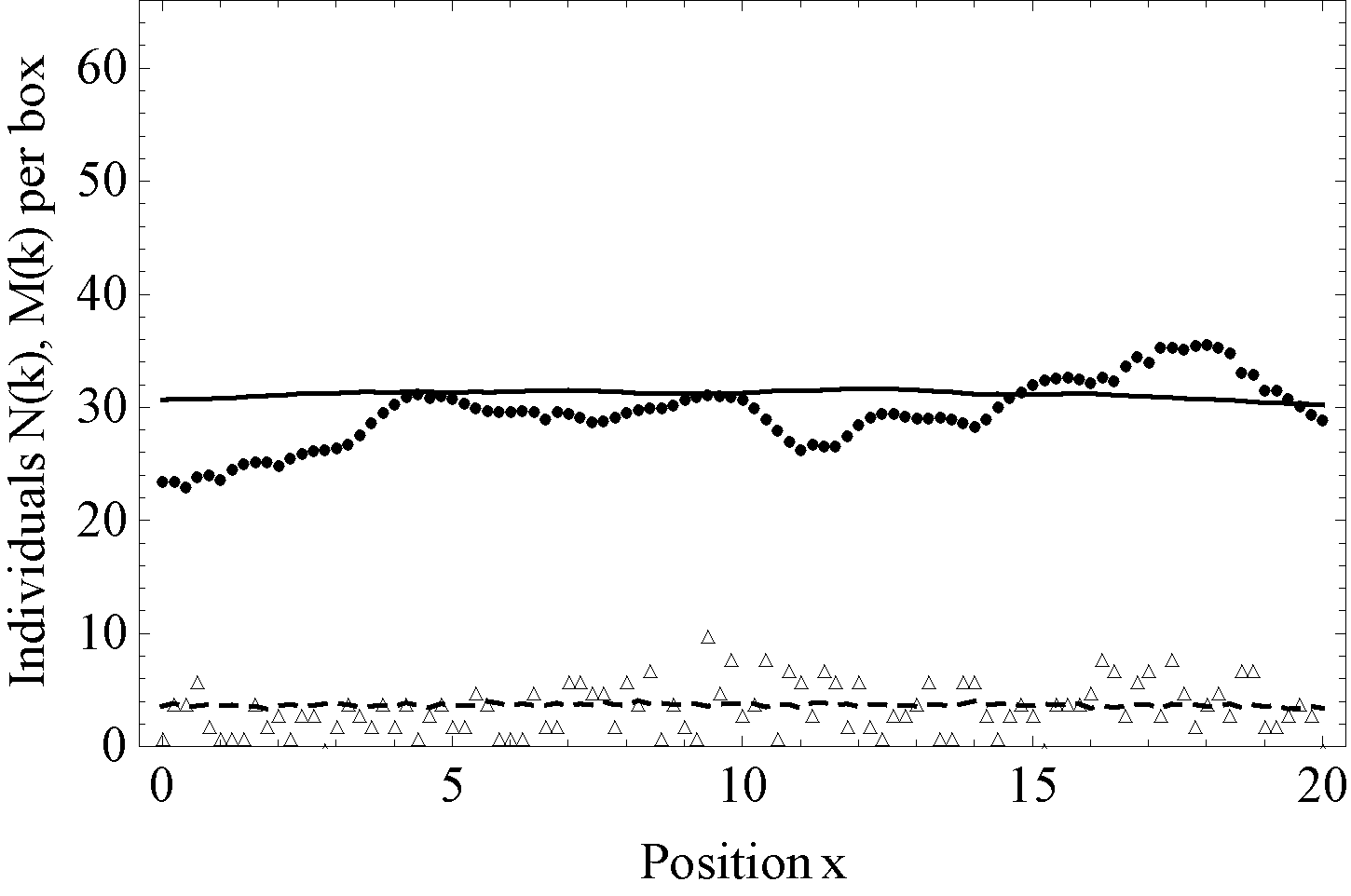}
}
\subfloat[Time Evolution Hybrid $\Theta=10$]
{
\label{fig:flatOscillationHybrid10_2}
 \includegraphics[width=0.48\linewidth]{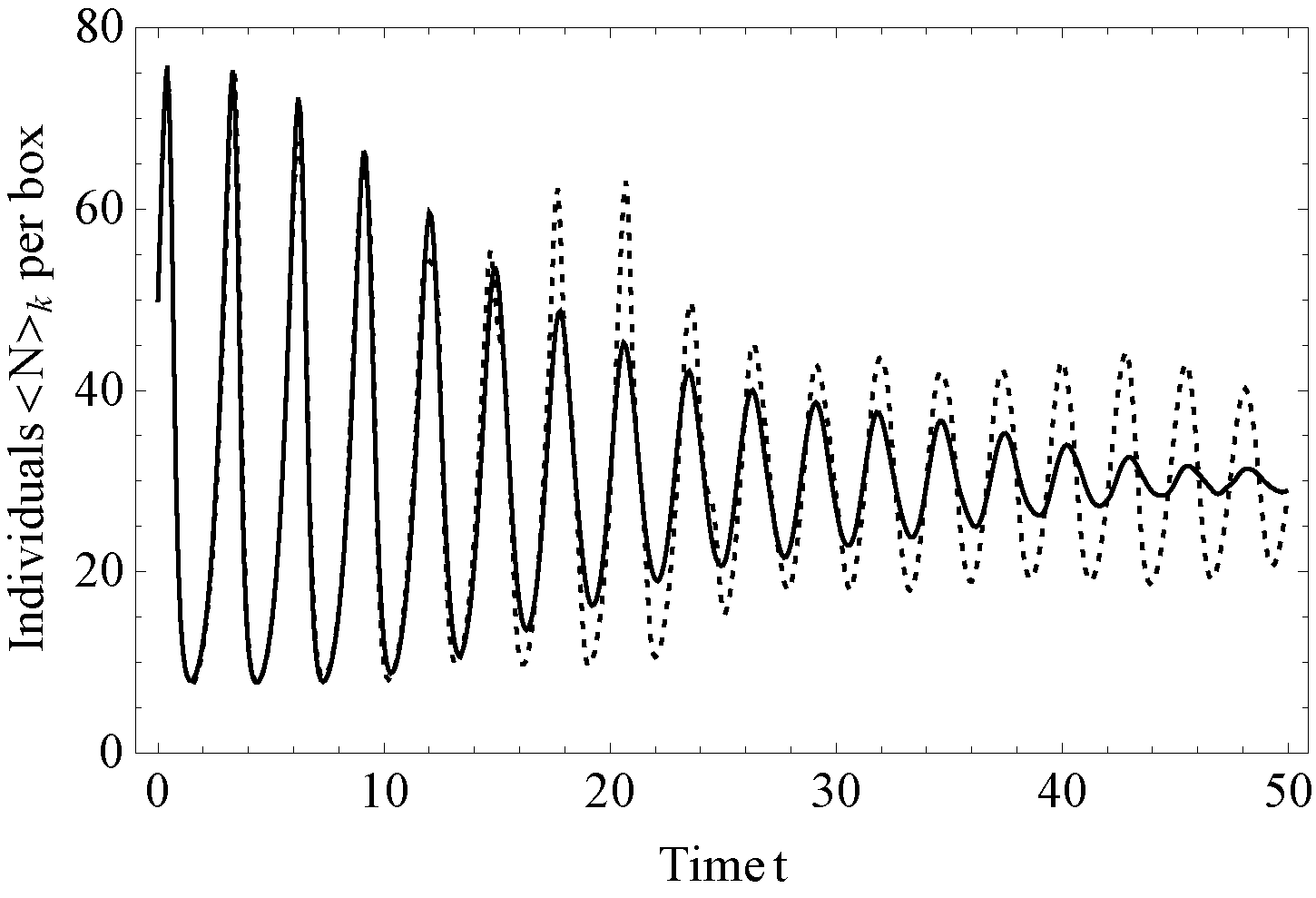}  

}\\
\subfloat[Spatial Profile Hybrid $\Theta=25$]
{
\label{fig:flatOscillationHybrid25_1}
\includegraphics[width=0.48\linewidth]{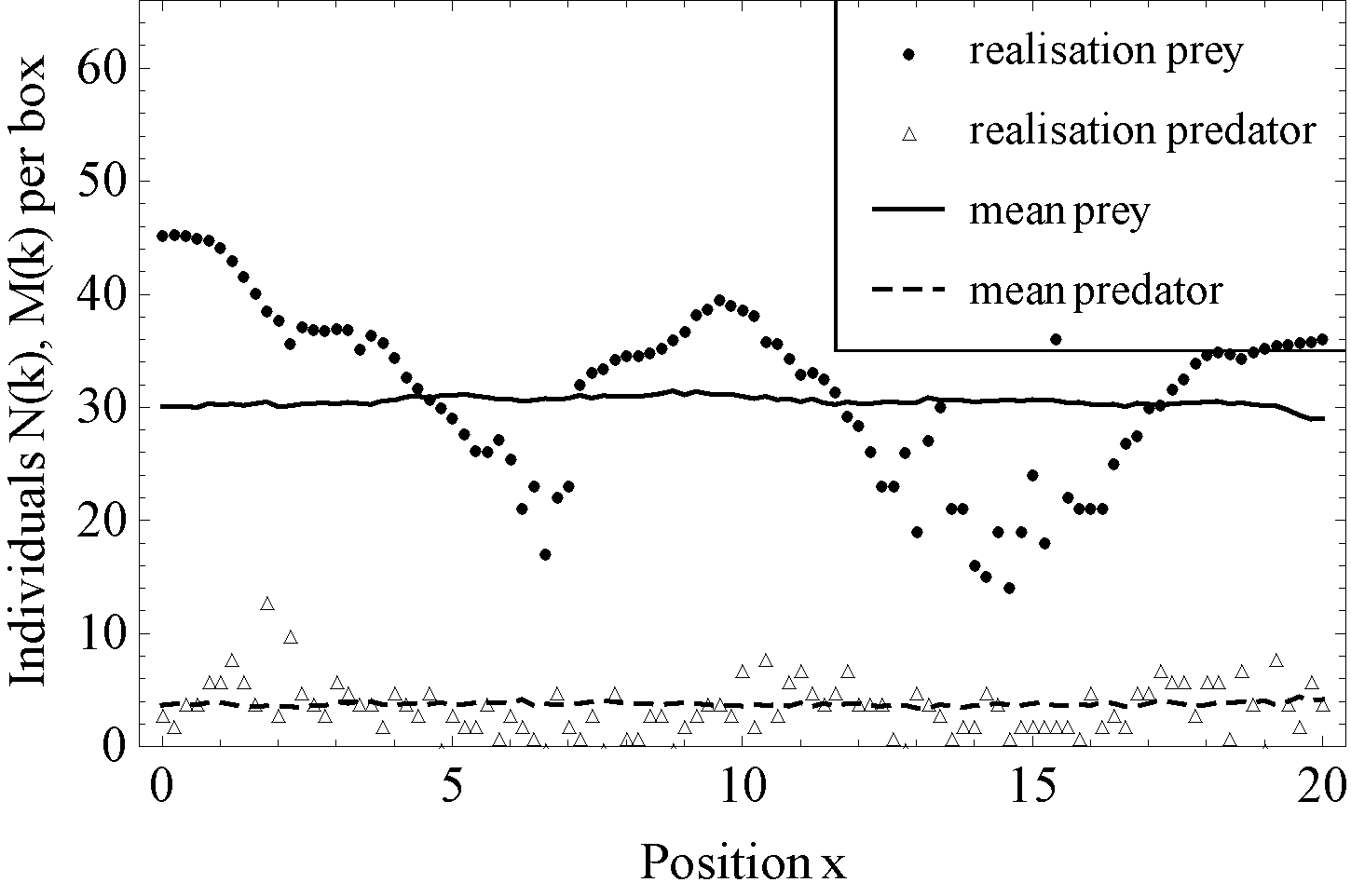}
}
\subfloat[Time Evolution Hybrid $\Theta=25$]
{
\label{fig:flatOscillationHybrid25_2}
\captionsetup[subfigure]{position=top}  
\includegraphics[width=0.48\linewidth]{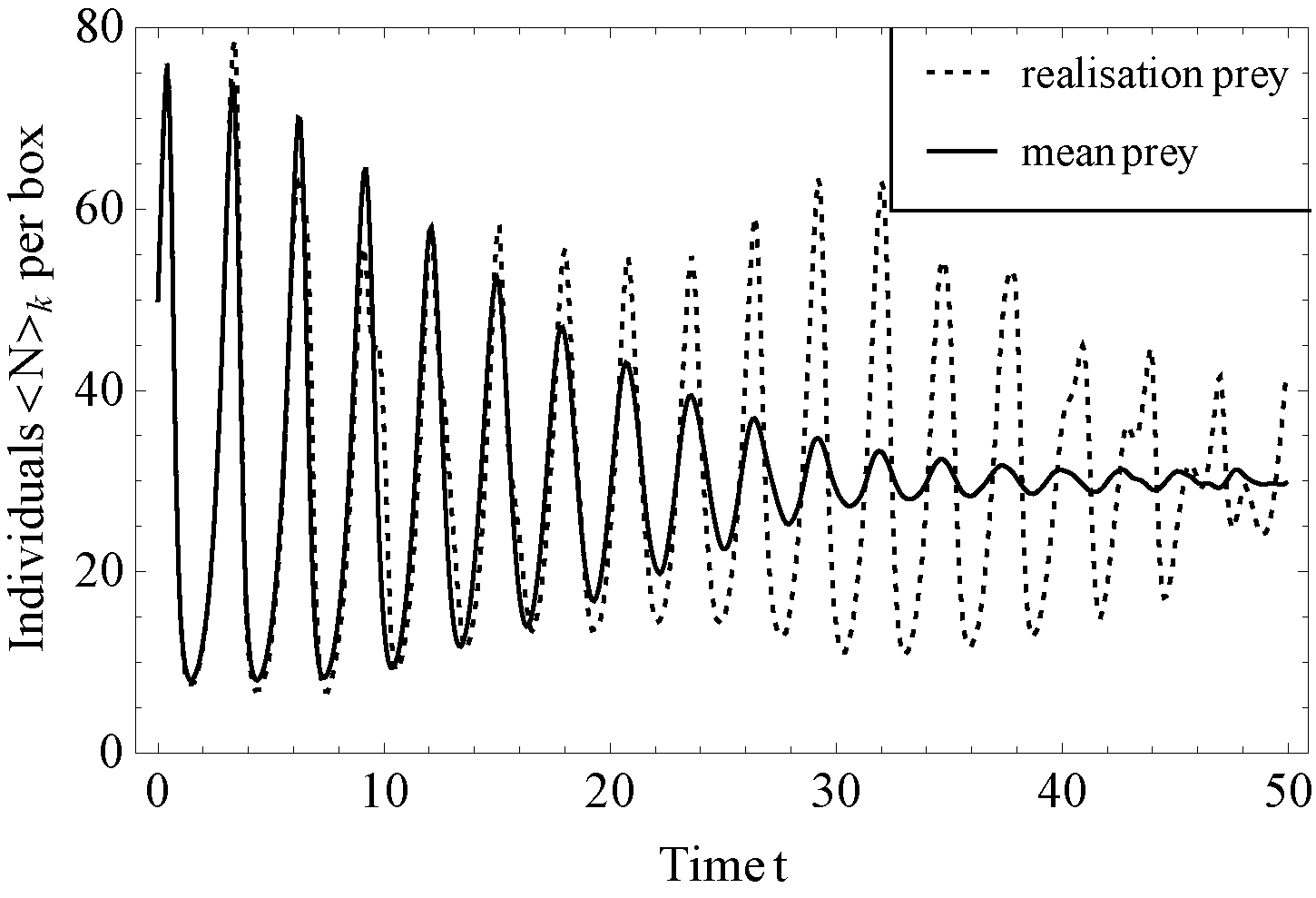}  

}
\caption{\label{fig:flatOscillation}Simulations of the spatial Lotka-Volterra Model with parameters $D_N=D_M=1$, $a=2,b=0.1,c=3$, $k_{max}=101, h=0.2$ and initial values $N(k,t=0)=50, M(k,t=0)=5$ for $k=1,\dots,k_{max}$. We compare \protect\subref{fig:flatOscillationStoch_1}-\protect\subref{fig:flatOscillationStoch_2}, the stochastic model, to \protect\subref{fig:flatOscillationHybrid10_1}-\protect\subref{fig:flatOscillationHybrid10_2}, the hybrid model with thresholds $\Theta=10$, and \protect\subref{fig:flatOscillationHybrid25_1}-\protect\subref{fig:flatOscillationHybrid25_2} with $\Theta=25$. The figures on the left show the spatial profile at time $t=4.1$ of the number of predators and prey of a single realisation as well as the mean of $256$ different realisations, whereas the figures on the right show the time evolution of the spatial average of numbers of prey in a single realisation and the mean of realisations. The corresponding PDE solution is shown in Figure \ref{fig:flatOscillation_LVPDE}.
}
\end{figure}
Corresponding results for the stochastic and hybrid models for two choices of the threshold values ($\Theta=10$ and $\Theta=25$) are shown in Figure \ref{fig:flatOscillation}. The column on the left shows the spatial profile of the number of predators and prey in a given box at time $t=2.5$, both for a single realisation and the mean of $256$ different realisations. Comparing either predator and prey numbers, we note that the means for the stochastic and hybrid models appear similar for both values of $\Theta$ (see Figure \protect\subref*{fig:flatOscillationStoch_1}\subref*{fig:flatOscillationHybrid10_1},\protect\subref{fig:flatOscillationHybrid25_1}), and are fairly homogeneous, whereas single realisations of either model differ markedly. As expected, the profile of predator and prey numbers in the stochastic model is noisy throughout the domain, whereas noise is suppressed in the hybrid model when the population numbers exceed the threshold. We observe the predator numbers in Figure \protect\subref*{fig:flatOscillationHybrid10_1}, are above the threshold and hence smoothly distributed in space, but they are not homogeneous, in constrast to the profile of the mean.

The column to the right in Figure \ref{fig:flatOscillation} shows the time evolution of the spatial mean of the number of prey, $\left\langle N\right\rangle_k=\frac{1}{k_{max}}\sum_{k=1}^{k_{max}}N(k)$. We note that the means for the stochastic model (Figure \protect\subref*{fig:flatOscillationStoch_2}) and the hybrid model are similar for either threshold (Figures \protect\subref*{fig:flatOscillationHybrid10_2},\protect\subref{fig:flatOscillationHybrid25_2}). In all cases, we observe oscillatory behaviour around $30=\frac{c}{b}$, the steady state of the corresponding PDE Eq. \eqref{eq:LVPDE}, with a decreasing amplitude. This contrasts with the dynamics of the PDE (see Figure \ref{fig:flatOscillation_LVPDE}) where the amplitude of oscillations was constant in time. This is because stochastic fluctuations may cause oscillations to fall out of synchrony and hence oscillations average out. Hence, the damping effect is stronger when plotting the mean of all realisations, rather than a single realisation, where the amplitude can fluctuate in time. However, this also implies that the PDE is not a good approximation to the mean over different realisations. We finally remark that while the plots of the spatial mean number of predators look different for the stochastic and hybrid models with the two thresholds, this is not significant as different realisations of the same model (not shown here) also look different. To properly compare the stochastic and hybrid models, we need quantitative measures (Table \ref{tab:LVoscillations_statisticalMeasures}).
\begin{table}[h]
\begin{tabular}{|c|c|c|c|c|}\hline
 					&  Stochastic & Hybrid $\Theta=25$& Hybrid $\Theta=10$ &Mean Field \\ \hline
$\mean{\left\langle {N}\right\rangle}_{k,t}$&	$30.50\pm 0.05$ & $30.32\pm 0.05$ & $30.30\pm 0.16$ &$30.47$\\
$\mean{\left\langle {M}\right\rangle}_{k,t}$&	$20.35\pm 0.04$ & $20.23\pm 0.04$ & $20.22\pm 0.13$ &$20.20$\\
$\sqrt{\mean{\left\langle {N}\right\rangle^2_{k,t}}-\mean{\left\langle {N}\right\rangle}_{k,t}^2}$ &	$16.0 \pm 0.3$ & $16.0\pm 0.3$ & $15.4\pm 0.4$ &\\
$\sqrt{\mean{\left\langle {M}\right\rangle^2_{k,t}}-\mean{\left\langle {M}\right\rangle}_{k,t}^2}$ &	$12.8 \pm 0.3$ & $12.8\pm 0.3$ & $12.3\pm 0.4$ &\\\hline
 \end{tabular}
  \caption{Mean of prey population over space, time and $256$ different realisations, 
  Eq. \eqref{mean_average}, as well as the corresponding number of predators. The standard deviation $\sqrt{\mean{\left\langle {N}\right\rangle^2_{k,t}}-\mean{\left\langle {N}\right\rangle}_{k,t}^2}=\sqrt{\frac{1}{256}\sum_{r=1}^{256}\left\langle {N_r}\right\rangle}$ of prey numbers, and likewise for predator numbers, are also shown, as are results for the corresponding mean field model, where there are no different realisations, and, hence, no standard deviation.}
 \label{tab:LVoscillations_statisticalMeasures}
 \end{table}
We calculare the spatial and temporal average of the mean number of prey across $256$ different realisations, 
\beq
\mean{\left\langle {N}\right\rangle}_{k,t}=\frac{1}{256 k_{max} t_{max}}\sum_{r=1}^{256}\sum_{k=1}^{k_{max}}\int_0^{t_{max}} N_r(k,t)dt,\label{mean_average}
\eeq
over the simulation time of $t_{max}=50$, and likewise the average of the mean number of predators. In this way we obtain a single number with large statistical significance, which is easier to compare as at individual time points, the oscillations in different realizations can be out of synchrony. We observe good agreement between the models. The values reported in Table \ref{tab:LVoscillations_statisticalMeasures} are in good agreement with the steady state values of the corresponding ODE model, but slightly different, as the temporal oscillations are not necessarily symmetric with respect to the steady state value. We then measure the standard deviation of the spatio-temporal averages, which is $\sqrt{\mean{\left\langle {N}\right\rangle^2_{k,t}}-\mean{\left\langle {N}\right\rangle}_{k,t}^2}$ for prey numbers, and likewise for predator numbers. Here, we note that the stochastic model is in close agreement with the hybrid model with $\Theta=25$; this agreement is less striking when $\Theta=10$. We conclude that the hybrid model can reproduce stochastic measures of the stochastic model, such as the standard deviation, and the agreement is better for larger values of the threshold $\Theta$.

\subsection{Extinction and Blow-up}\label{sec:extinctionBlowup}

\begin{figure}[h!]
\subfloat{
\includegraphics[width=0.48\linewidth]{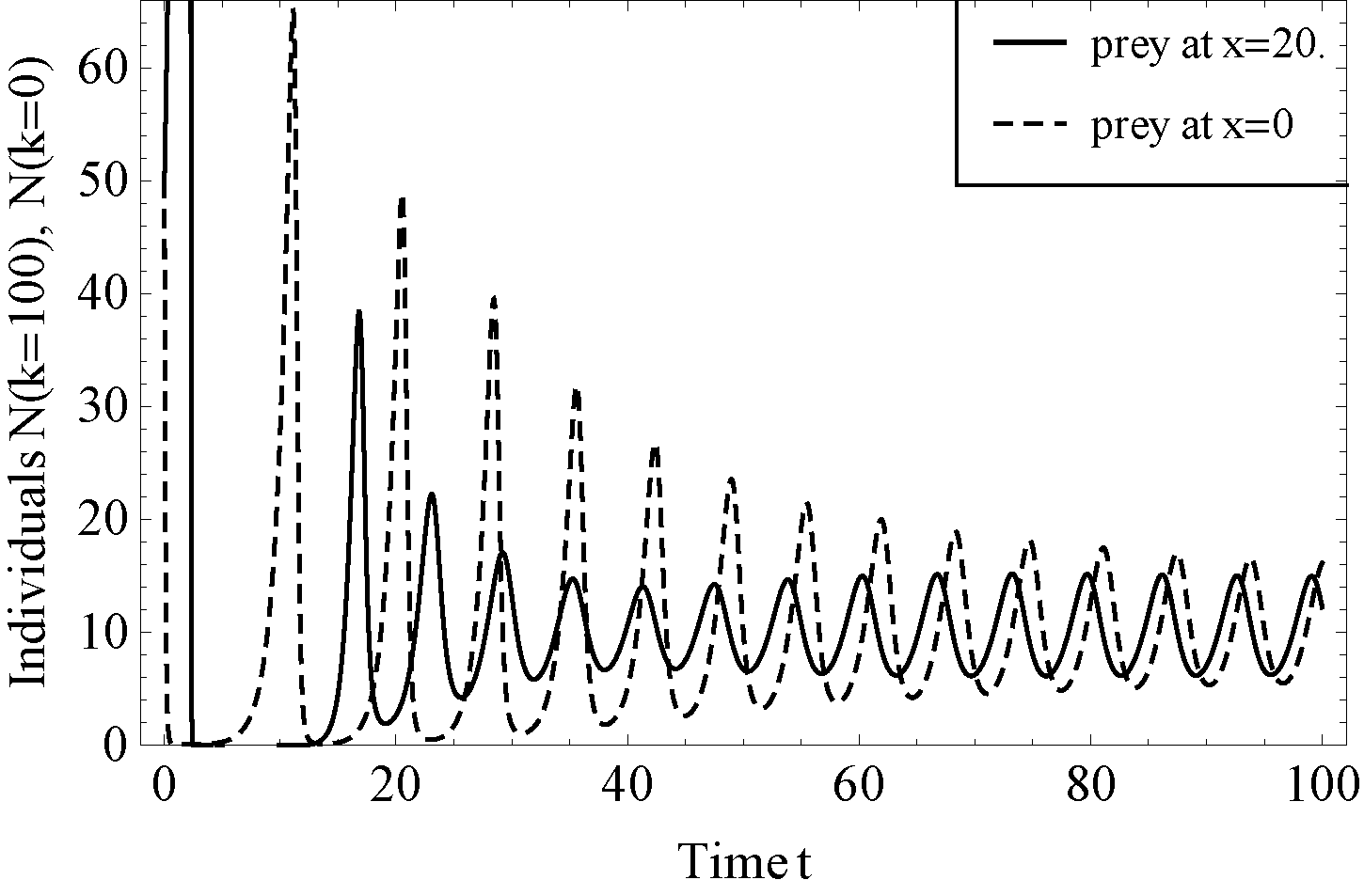}
}
\subfloat{
\includegraphics[width=0.48\linewidth]{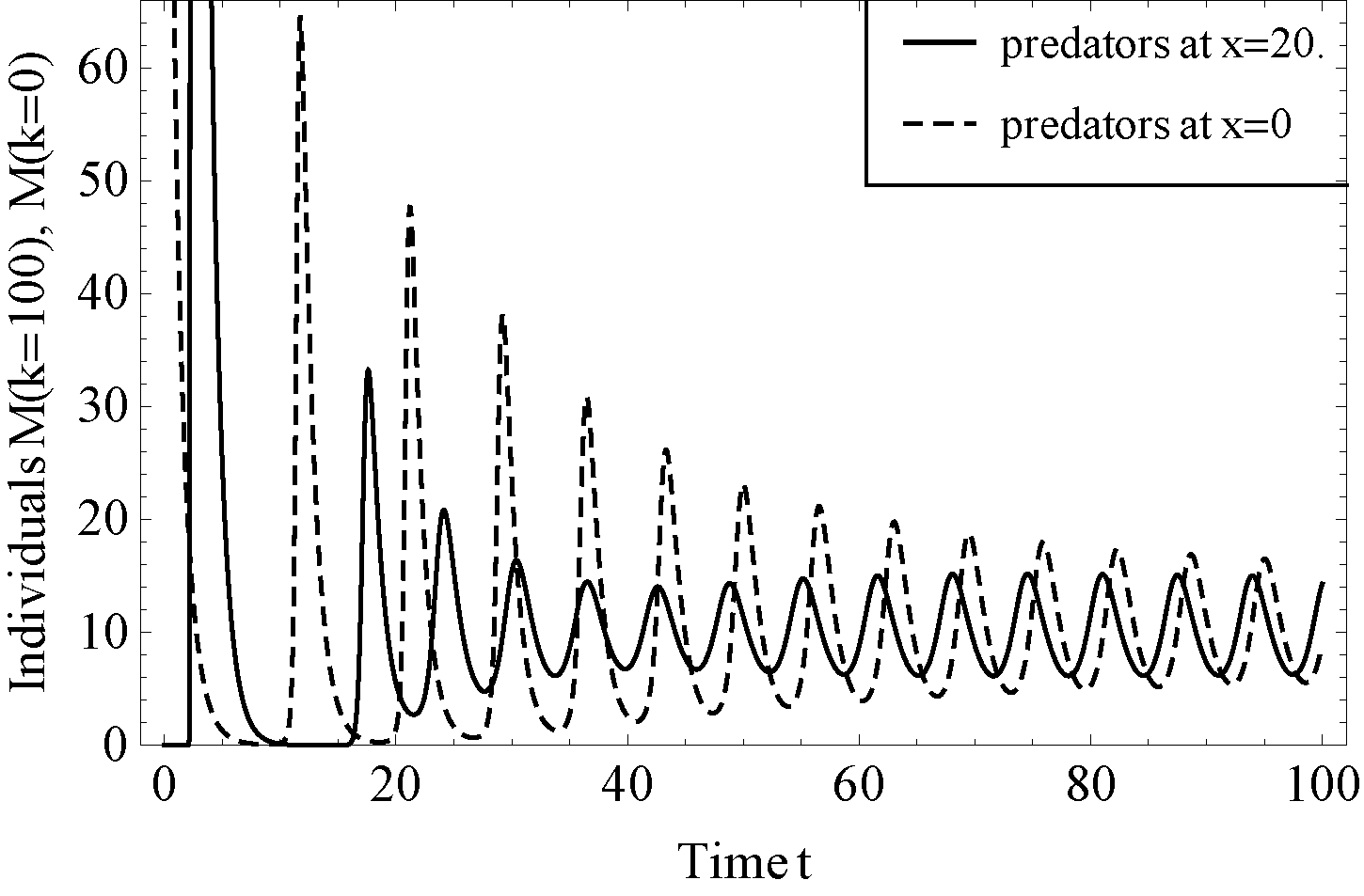}
}
\caption{\label{fig:extinction_PDE_two_locations}Number of prey, left, and predators, right, obtained from the deterministic model, equation \eqref{eq:LVPDE}, at the two boundaries, $x=0$ and $x=20$, as a function of time. The initial steep rise of the number of prey at $x=20$ is due to the absence of predators in this region and cannot be fully seen here. It is followed by a sharp peak of predators. We see that during the first few oscillations the number of individuals can often be close to zero, indicating that extinction would be possible at those times in the corresponding stochastic model. At later times $t>20$ the number of prey and predators is fluctuating with minima sufficiently away from zero, making extinction at those times less likely.}
\end{figure}

We now choose a spatially homogeneous population of prey, $N(k,t=0)=50$, $k=1,\dots,k_{max}$, a number of predators present only on the left side of the domain, $M(k)=100, (k=1,\dots,9)$, $M(10)=98$, $M(11)=50$, $M(12)=2$, and $M(k)=0, (k=13,\dots 101)$, Neumann boundary conditions and parameters $D_N=D_M=1$, $a=1,b=0.1,c=2$, $L=20, k_{max}=101$. This scenario is typical example of invasion of a predator into a population of prey, leading to spatial and temporal fluctuations even in the purely deterministic model (Figure \ref{fig:extinction_PDE_two_locations}).

Figure \ref{fig:extinction_PDE_two_locations} shows how the number of predators and prey at the two boundaries ($x=0$, $x=20$) vary time for the mean field model. We note that during the first few oscillations both predator and prey populations are close to zero. 
In the corresponding stochastic model, the population can only be integer-valued, so a value below $1$ in the deterministic model indicates that extinction of the population is likely. At later times, we observe regular oscillations with minima significantly above $0$, and conclude that if extinction were to occur in the stochastic model, it would most likely happen at early times.
\begin{figure}[h!]
\subfloat[Stochastic Model, Oscillation]
{
\label{fig:extinction_stoch_mean_osc}
\includegraphics[width=0.48\linewidth]{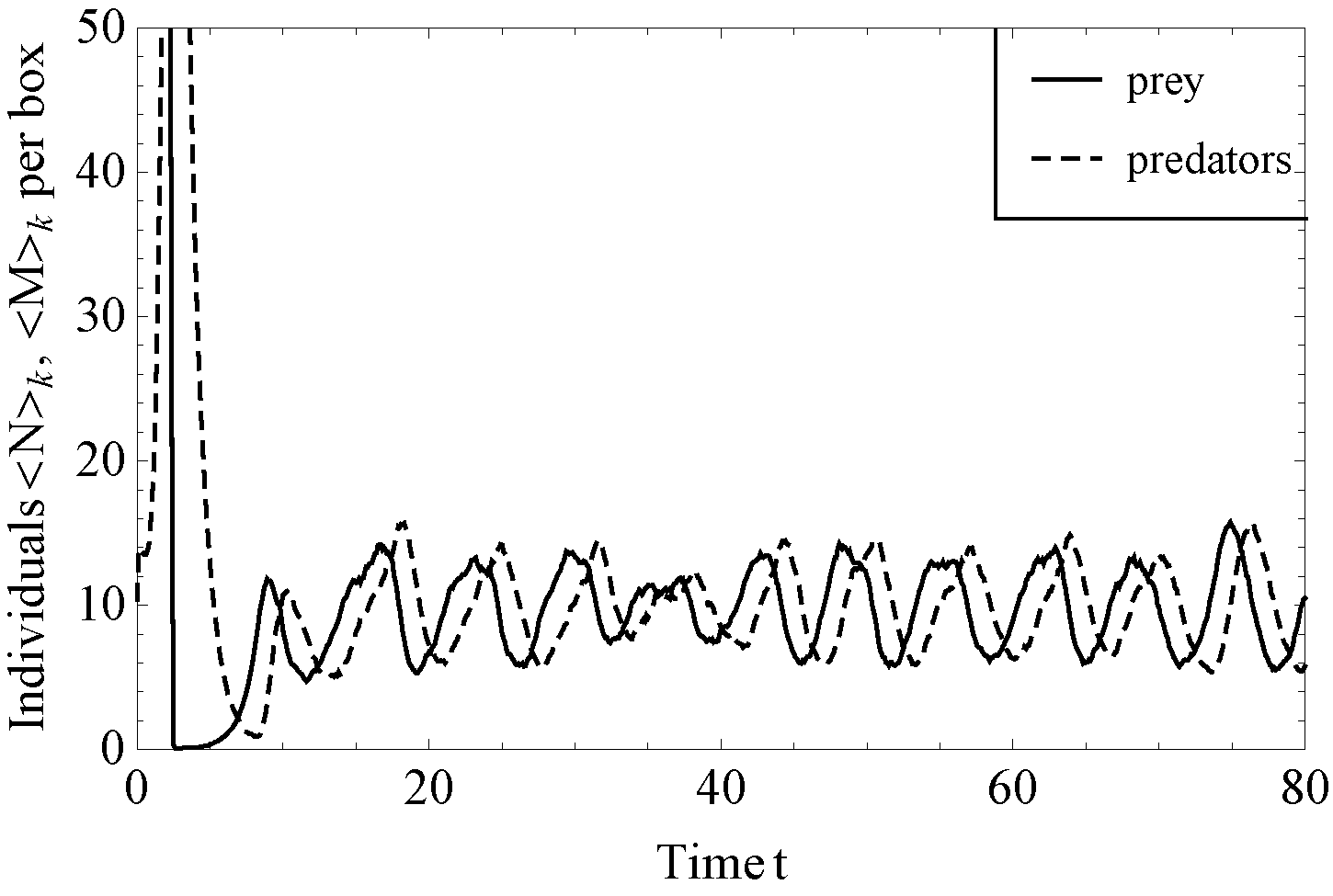}
}
\subfloat[Hybrid Model, Oscillation]
{
\label{fig:extinction_hybrid_10_mean_osc}
\includegraphics[width=0.48\linewidth]{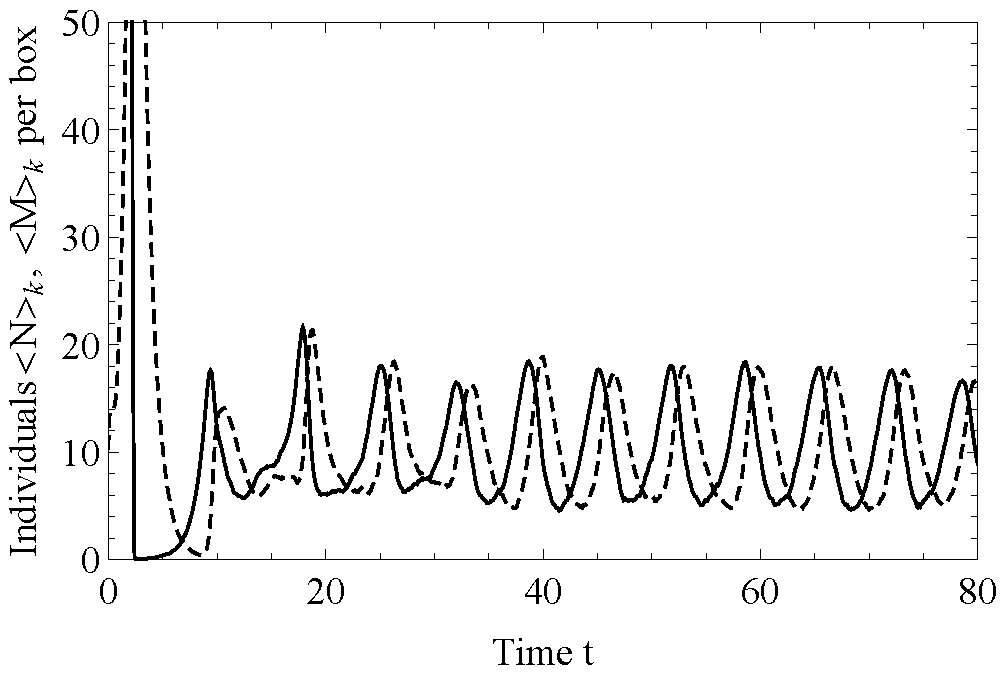}
}\\
\subfloat[Stochastic Model, Extinction]
{
\label{fig:extinction_stoch_mean_ext}
\includegraphics[width=0.48\linewidth]{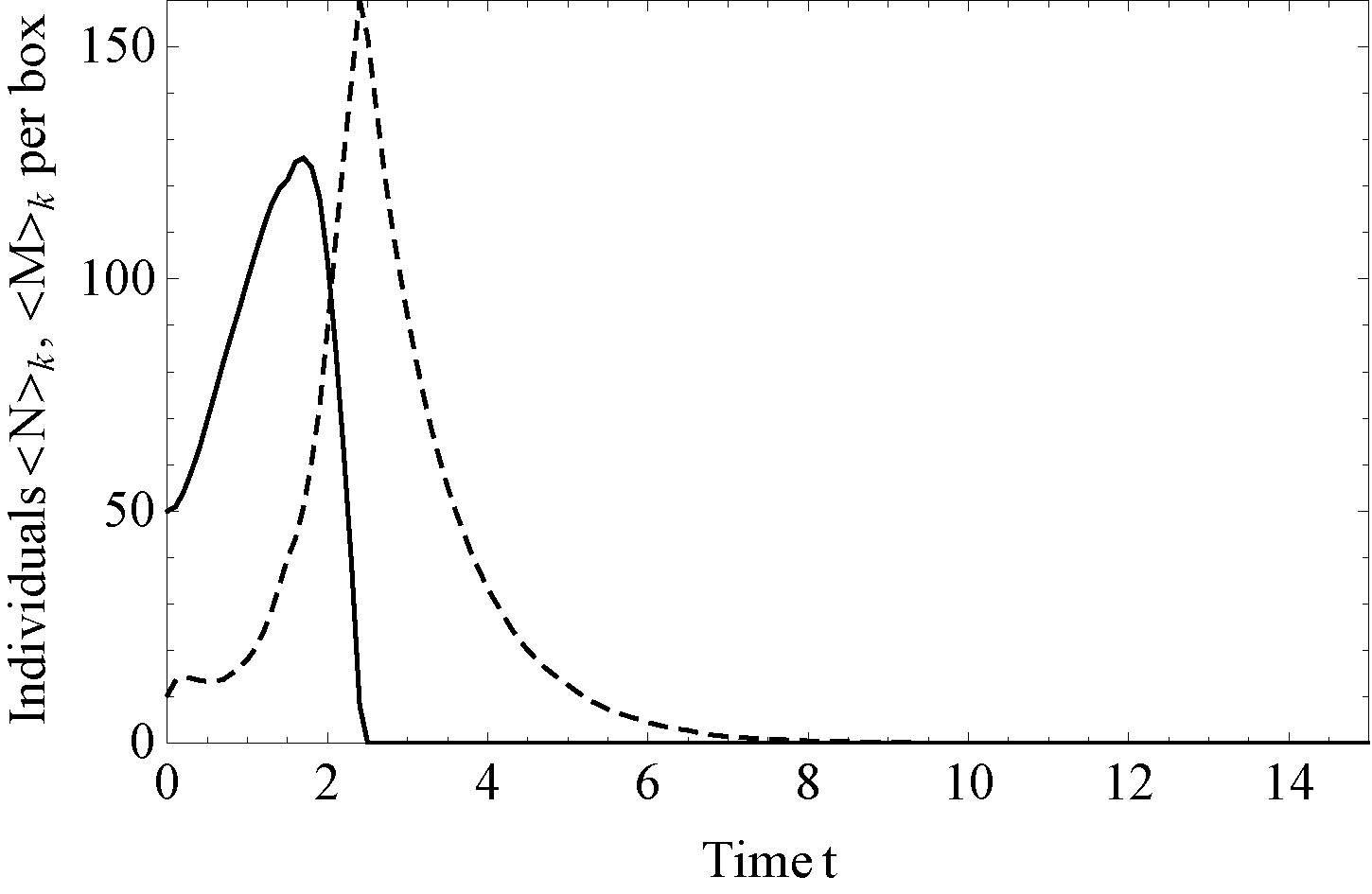}
}
\subfloat[Hybrid Model, Extinction]
{
\label{fig:extinction_hybrid_10_mean_ext}
 \includegraphics[width=0.48\linewidth]{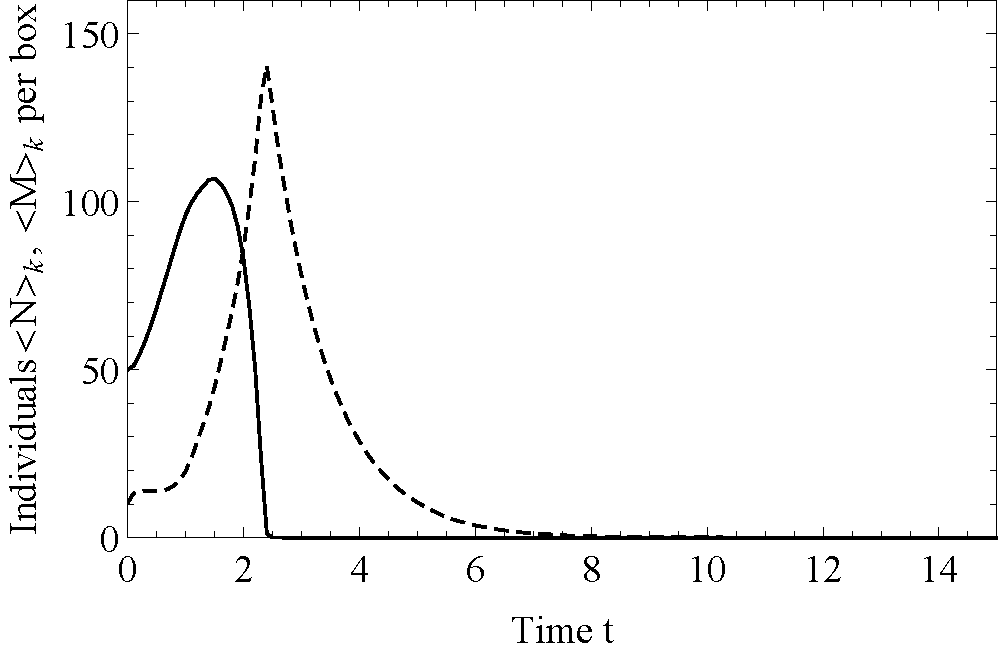}  

}\\
\subfloat[Stochastic Model, Blow-Up]
{
\label{fig:extinction_stoch_mean_blow}
\includegraphics[width=0.48\linewidth]{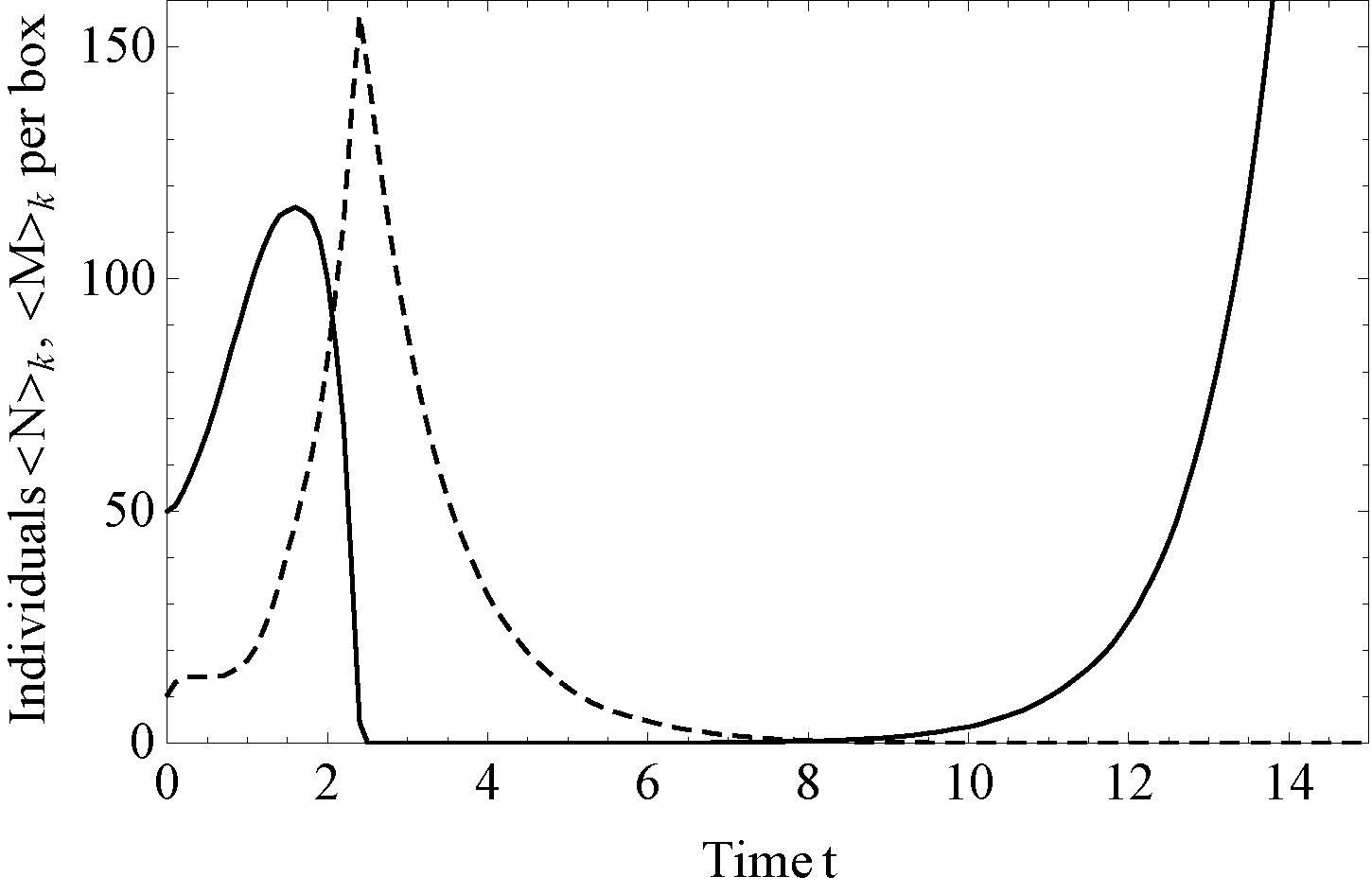}
}
\subfloat[Hybrid Model, Blow-Up]
{
\label{fig:extinction_hybrid_10_mean_blow}
\includegraphics[width=0.48\linewidth]{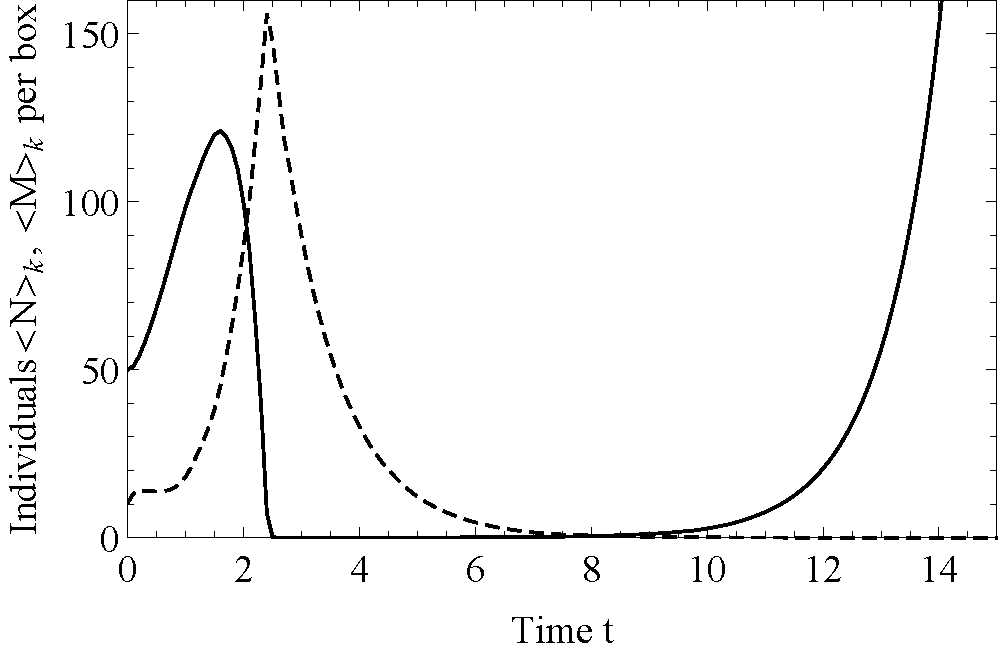}  
}
\caption{\label{fig:LV_extinction_hybrid_mean} Simulations of the spatial Lotka-Volterra Model with parameters $D_N=D_M=1$, $a=1,b=0.1,c=1$, $L=20, k_{max}=101$. We plot the spatial average $\left\langle {N}\right\rangle_{k}=\frac{1}{k_{max}}\sum_{k=1}^{k_{max}}N(k)$ of the number of predators, and likewise prey, in a box over time for three realisations of the stochastic model, \protect\subref{fig:extinction_stoch_mean_osc},\protect\subref{fig:extinction_stoch_mean_ext},\protect\subref{fig:extinction_stoch_mean_blow}, and three realisations of the hybrid model, \protect\subref{fig:extinction_hybrid_10_mean_osc},\protect\subref{fig:extinction_hybrid_10_mean_ext},\protect\subref{fig:extinction_hybrid_10_mean_blow}, with a threshold of $\Theta=10$. These three realisations of each model show the three qualitatively different outcomes, namely, oscillatory solutions, extinction of both species or extinction of predators and subsequent blow-up of prey.
}
\end{figure}
Figure \ref{fig:LV_extinction_hybrid_mean} confirms these expectations. Figure \subref*{fig:extinction_stoch_mean_osc} shows that, for the stochastic model, the spatial average of the number of predators (or prey) may exhibit oscillations similar to those of the deterministic model. However, extinction of either population can also occur. Figure \subref*{fig:extinction_stoch_mean_ext} shows that if the prey die out first, then, necessarily, the predators will also die out as well. On the other hand, if the predators die out first, then prey numbers will blow-up (see Figure \subref*{fig:extinction_stoch_mean_blow}). 
Figures \subref*{fig:extinction_hybrid_10_mean_osc},\subref{fig:extinction_hybrid_10_mean_ext},\subref{fig:extinction_hybrid_10_mean_blow} confirm that the hybrid model can reproduce each of these three qualitatively different scenarios. Since statistics for the mean and standard deviation are not meaningful if the prey population blows up, we compare the frequency of extinction events in the stochastic and hybrid models. 
\begin{table}[h]
 \begin{tabular}{c|c|c|c|c|c}
 &  Stochastic & Hybrid 100& Hybrid 50 & Hybrid 25 & Hybrid 10\\ \hline
 Prey & 0.46 & 0.45 & 0.46 & 0.45 & 0.43 \\ 
 Predator & 0.59 & 0.57 & 0.59 & 0.61 & 0.56 \\
 \end{tabular}
 \caption{Shown is the extinction probability for the prey and predator population in the Lotka-Volterra system for the scenario as described in this section. This probability is calculated by repeating $256$ simulations for each of the stochastic model and the hybrid model with thresholds of $10,25,50$ and $100$, and recording at time $t=80$ if a population is extinct or not.}
 \label{tab:extinction}
\end{table}

The results presented in Table \ref{tab:extinction} suggest that the hybrid model with a threshold of $\Theta=10$ or larger has a similar probability of extinction as the stochastic model. We conclude that in this case the hybrid model provides a good approximation to the stochastic model. We finally investigate the performance gain obtained by using the hybrid model.

\begin{table}[h]
 \begin{tabular}{|c|c|c|c|}
 \hline
 &  Stochastic & Hybrid 100& Hybrid 10   \\ \hline
 Simulation time in hours & 40.0 & 10.3 & 4.8  \\ \hline
 \end{tabular}
	\caption{Performance evaluation of hybrid model. We performed 256 simulations for each of the pure stochastic model as well as the hybrid model with a threshold of $10$ and $100$. As expected, we see a significant performance gain when using the hybrid model compared to the stochastic model, and the speedup is larger the lower the threshold is, as then we switch earlier to the PDE. 
	}
 \label{tab:performance}
\end{table}
Table \ref{tab:performance} compares the computational time needed to perform 256 simulations of the stochastic and hybrid models proceeding until extinction or time $t=100$ in all cases. The simulations were performed on a Xeon-2680 16-core server with 2.7 GHz and 128GB RAM. Simulations of the stochastic model were stopped once predator extinction occurred, as in that case the population of prey necessarily blows up. Hence, the advantage in performance of the hybrid model is even larger than the numbers indicate. 

\section{Conclusions}\label{sec:conclusions}

In this paper we have presented a hybrid algorithm which couples a stochastic reaction-diffusion system on a lattice to its associated mean field limit, which can be seen as the finite-difference discretisation of a reaction-diffusion PDE. Our algorithm preserves mass at the interface between the stochastic and deterministic domains, and these domains need neither to be static nor connected. Furthermore, for multi-species systems, the corresponding stochastic and deterministic domains may differ for individual species. We also introduced a normalisation procedure to ensure that the stochastic domain contains only integer numbers of particles when the mean field equations are evolved or the interface is moved. With the example of the spatial Lotka-Volterra system, this paper provides a detailed study of a multi-species hybrid system that can accommodate multiple moving interfaces. We found that our hybrid algorithm can produce stochastic effects that are not present in the corresponding mean field model with the same frequency as the stochastic model, while the time taken to perform hybrid simulations is much shorter than that for fully stochastic simulations.

At present, we solve the deterministic equations on the same grid as the stochastic model. For computational reasons, one might consider other numerical schemes to solve the mean field equations \cite{engblom2009,isaacson2006incorporating}. This could be particularly important for simulating systems in higher spatial dimensions and geometries with curved boundaries, in such cases, finite difference discretisations might not be sufficiently accurate. Another modification to determinate the position of our algorithm could be to refine the condition used to the interface. For the models studied in this paper, we found that counting the number of particles in a box provides a good threshold condition for when to use the stochastic, and when to use the mean field, model. This is justified as typically stochastic fluctuations scale with the square root of the number of particles. However, for some systems one might need to choose the domains directly according to the size of the fluctuations.

Finally, it would be interesting to test our algorithm on reaction-diffusion systems involving larger numbers of individuals, such as models of cancer growth \cite{gatenby1996reaction,swanson2003virtual}, angiogenesis \cite{spill2014mesoscopic} or cell polarity \cite{mckane2014stochastic}. Such considerations are beyond the scope of this paper and are therefore postponed for future research.

\section{Acknowledgements}
This publication was based on work supported in part by Award No KUK-C1-013-04, made by King Abdullah University of Science and Technology (KAUST). TA gratefully acknowledges the Spanish Ministry for Science and Innovation (MICINN) for funding under grants MTM2011-29342 and Generalitat de Catalunya for funding under grant 2009SGR345. PG acknowledges Wellcome Trust [WT098325MA] and Junta de Andalucía Project FQM 954.

\appendix

\section{Alterations of the Hybrid Algorithm}\label{app:alterationsAlgorithm}

\subsection{Large Stochastic Time Step}\label{app:largeTimeStep}

A significat problem with continuous diffusion equations is that, in finite times, mass can spread arbitrarily far. As a result when the PDE is evolved, its solution can  leak arbitrarily far into the stochastic domain unless this is prevented by the imposittion of an artificial condition at the interface. In our hybrid algorithm, this problem is avoided by using a finite-difference approximation of the PDE with the same lattice size as the stochastic model so that on each time step the PDE solution spreads by only one spatial compartment. Thus, the PDE solution can, in one time step, enter the interface region, but not the stochastic domain. We remark that the infinitely fast spread associated with the continuous diffusion equation is not physically realistic, it consequently the finite difference approximation to the diffusion PDE is not necessarily less realistic than the  PDE itself.

A problem can arise when the time $\tau$ to the next event is larger than the maximum time step $\tau_{PDE}$ such that the finite difference scheme converges. In this case, we need several iterations of \eqref{eq:finiteDifferenceGeneralEuler} to evolve the PDE until time $\tau$, and during this time the PDE solution could leak into the stochastic domain. This problem can be avoided by increasing the size of the interface region, or one could choose a smaller lattice constant for the PDE regime (compared to the stochastic domain). In the examples shown in the present paper, the problem of $\tau>\tau_{PDE}$ is rare. When we encountered $\tau>\tau_{PDE}$, we chose $i$ such that the finite difference scheme \eqref{eq:finiteDifferenceGeneralEuler} converges with time step $\frac{\tau}{i}$. We then iterated \eqref{eq:finiteDifferenceGeneralEuler} $i$ times with time step $\frac{\tau}{i}$, keeping the interface, still consisting of one compartment only, fixed. Hence, the interface is treated as a Neumann-no-flux boundary during the time step $\tau$. We observed that the total changes of mass in the interface box is small relative to the total amount of mass present, so the error we introduce due to the artificial Neumann condition is negligible.

A related problem is that when mass moves from the PDE into the interface domain, this, in principle, changes the transition rates of stochastic reactions of the interface, and drift terms to the master must be added equation. However, as discussed above, for the simulations performed in this paper the change of mass in the interface compartment is small within one time step. Hence, the change in the transition rates one will also be is small and can be neglected.

\subsection{Small Stochastic Time Step}\label{app:smallTimeStep}
If the random Gillespie time step is much smaller than the PDE convergence time step, $\tau\ll\tau_{PDE}$, a simple modification of our algorithm consists of not iterating the mean field domain during every time step, but only after $I$ time steps such that for the $i$ Gillespie time steps $\tau_i$, we have $\sum_{i=1}^I\tau_i <\tau_{PDE}$, but the next time step is likely to bring the cumulative time step above $\tau_{PDE}$.
\subsection{Deterministic Interface Reactions}\label{app:deterministicInterfaceReaction}

The interface is chosen such that the mean-field equations are sufficiently accurate to represent the system at the interface. In section \ref{sec:interfaceConditionGeneral}, we have described the reactions in the interface compartment in a stochastic way, but we could also describe the reactions deterministically by replacing equation \eqref{eq:finiteDifferenceInterfaceFlux} with
\beq
\label{eq:finiteDifferenceInterfaceFluxReactions}
n^s_{k_{I}}(t+\tau) = n^s_{k_{I}}(t) + \tau \left(\frac{D^s}{h^2}\left(n^s_{k_I}(t)-n^s_{k_{I}-1}(t)\right) + r^s(n^1_{k_I}(t),\dots,n^{s_{max}}_{k_I}(t))\right).
\eeq
Then, the transition rates describing stochastic reactions in the interface compartment should be set to zero.

\subsection{Multiple Interfaces for Different Species}\label{app:multipleInterfaces}

In many situations, the regions where the concentration of one species is high, and hence the deterministic PDE description is valid, are different for different species. Hence, there is often a requirement to have separate interfaces for the different species. We can use the interface condition as in section \ref{sec:interfaceConditionGeneral} for each species separately. However, we will now encounter regions of space $[L_{I_1},L_{I_2}]$ where some species are modelled stochastically and others deterministically. Let $N_1(k)$ be stochastic for $k=k_{I_1},\dots,k_{I_2}$, and $n_2(k)$ be deterministic in the same interval, where, as before, we identify $k_{I_1} h = L_{I_1}$, $k_{I_2} h = L_{I_2}$. For simplicity, we focus on a single reaction involving only those two species, which in the full stochastic model would be written as
\beq
\mathcal{T}_{N_1(k)+\rho_1,N_2(k)+\rho_{2}|N_1(k),N_2(k)} = R(N_1(k),N_2(k)).
\eeq
As species $2$ is modelled deterministically, we replace $N_2(k)$ by $h n_2(k)$ and obtain
\beq
\label{eq:transitionRateMixedSpecies}
\mathcal{T}_{N_1(k)+\rho_1,h n_2(k)+\rho_{2}|N_1(k),hn_2(k)} = R(N_1(k),h n_2(k)).
\eeq
Hence, the reaction is still stochastic. However, $hn_2(k)$ is real valued. The situation is thus similar to how reactions were dealt with at a single-species interface. The real-valuedness does not cause any problems as we assume $hn_2(k)$ is large in the interface region. If the reaction vector $\rho_{2}$ is negative, we have to assume $hn_2(k)\gg \rho^r_{2}$, so that a single stochastic reaction cannot lead to negative values of $n_2(k)$. This is easily ensured by choosing the hybrid model threshold accordingly high.

We can regard this reaction as a stochastic source for an otherwise deterministic $n_2(k)$. The deterministic part of the evolution of $n_2(k)$ is described by
\begin{align}
\label{eq:finiteDifferenceMixedSpecies}
 n_2(k,t+\Delta t) &= n_2(k,t) \nonumber\\
 &+ \Delta t\left(\frac{D^2}{h^2}\left( n_2(k+1,t)+n_2({k-1},t)-2n_2(k,t)\right) + r_2\left(n_{2}(k)\right)\right),\nonumber\\
 k_{I_1}<k<k_{I_2}.
\end{align}
The corresponding equations at $k=k_{I_1}$ or $k=k_{I_2}$ require modifications of the flux term in the same way as the equations in the single species case, equation \eqref{eq:finiteDifferenceInterfaceFlux}. We have denoted reactions which only depend on $n_2(k)$ as $r_2\left(n_{2}(k)\right)$.

The finite difference equation \eqref{eq:finiteDifferenceGeneralEuler} was chosen such that it produces exactly the mean behaviour of the underlying stochastic model. We now confirm that the mixed model \eqref{eq:transitionRateMixedSpecies} and \eqref{eq:finiteDifferenceMixedSpecies} still produces the same mean behaviour in the limit of large particle numbers. This means we assume $N_1(k)\propto \Omega$, with $\Omega\gg1$, whereas $n_2(k)\propto \Omega$ already to justify the use of the deterministic equations. 
We have, by the derivation of \eqref{eq:generalMeanField}, that 
\begin{align}
\mean{\mathcal{T}_{N_{1}(k)+\rho^r_1,N_{2}(k)+\rho^r_{2}|N_{1}(k),N_{2}(k)}} &= \mean{R(N_{1}(k),N_{2}(k))} \nonumber\\
&= \Omega\left(r(n_1(k),n_{2}(k)) + \mathcal{O}\left(\frac{1}{\Omega^{1/2}}\right)\right).
\end{align}

But this means we will also have 
\beq
\mean{R\left(\frac{N_1(k)}{h},n_{2}(k)\right)} = \Omega\left(r\left(n_1(k),n_{2}(k)\right) + \mathcal{O}\left(\frac{1}{\Omega^{1/2}}\right)\right),
\eeq
as required.



  \bibliographystyle{elsarticle-harv} 
  \bibliography{stochastic}





\end{document}